\documentclass[aps,prb,twocolumn,showpacs,superscriptaddress]{revtex4-1}  % for review and submission
\usepackage{graphicx}  % needed for figures
\usepackage{dcolumn}   % needed for some tables
\usepackage{bm}        % for math
\usepackage{amssymb}   % for math
\usepackage{bbold}
\usepackage{braket}
\usepackage{amsmath}
\usepackage{todonotes}
\usepackage{siunitx}
\usepackage{chemformula}
\usepackage{subfigure}
\usepackage{tikz}
\usetikzlibrary{shapes,arrows, calc}
\usepackage[T1]{fontenc}
\newcommand{\beq}{\begin{equation}}
\newcommand{\eeq}{\end{equation}}
\newcommand{\eqname}{Eq.}
\newcommand{\secname}{Sec.}
\newcommand{\forcefield}{q-SPC/FW+anh}
\newcommand{\fscha}{{\mathcal F(\vec{\mathcal R}, \mathbf{\Phi}, \left\{\vec a_i\right\})}}
\newcommand{\fscham}{{\mathcal F(\vec{\mathcal R}, \left\{\vec a_i\right\})}}
\newcommand{\rscha}{{\rho_{\vec{\mathcal R}, \mathbf{\Phi}}}}
\newcommand{\rscham}{{\rho_{\vec{\mathcal R}}}}
\DeclareMathOperator{\Tr}{Tr}

\DeclareSIUnit\rydberg{Ry}

\begin{document}

% The following information is for internal review, please remove them for submission
\widetext
% \leftline{Version 1 as of \today}
% \leftline{Primary authors: Lorenzo Monacelli}
% \leftline{To be submitted to PRB}
% %\leftline{Comment to {\tt d0-run2eb-nnn@fnal.gov} by xxx, yyy}
% \centerline{\em D\O\ INTERNAL DOCUMENT -- NOT FOR PUBLIC DISTRIBUTION}

% the following line is for submission, including submission to the arXiv!!
%\hspace{5.2in} \mbox{Fermilab-Pub-04/xxx-E}

\title{Pressure and stress tensor of complex anharmonic crystals\\ within the stochastic self-consistent harmonic approximation}
% D0 authors (remove the first 3 lines
                             % of this file prior to submission, they
                             % contain a time stamp for the authorlist)
% (includes institutions and visitors)

\author{Lorenzo Monacelli}
\affiliation{Department of Physics, University of ``La Sapienza'', Rome, Italy}
\author{Ion Errea}
\affiliation{Fisika Aplikatua 1 Saila, Bilboko Ingeniaritza Eskola, University of the Basque Country (UPV/EHU), Bilbao, Basque Country, Spain}
\affiliation{Donostia International Physics Center (DIPC), Donostia-San Sebastian, Basque Country, Spain}
\author{Matteo Calandra}
\affiliation{Sorbonne Universit\'e, CNRS, Institut des Nanosciences de Paris, UMR7588, F-75252, Paris, France}
\author{Francesco Mauri}
\affiliation{Department of Physics, University of ``La Sapienza'', Rome, Italy}
\affiliation{Graphene Labs, Fondazione Istituto Italiano di Tecnologia, Genova, Italy}
\date{\today}

\begin{abstract}

The self-consisted harmonic approximation (SCHA) allows the computation of free energy of anharmonic crystals considering both
quantum and thermal fluctuations. Recently, a stochastic implementation of the SCHA has been developed, tailored for applications that use
 total energy and forces computed from first principles. In this work, we extend the applicability of the stochastic SCHA to complex crystals with many degrees of freedom, with the optimisation of both the lattice vectors and the atomic positions. To this goal, we provide an expression for the evaluation of the pressure and stress tensor within the stochastic SCHA formalism. Moreover, we develop a more robust free energy minimisation algorithm, which allows us to perform the SCHA variational minimisation  very efficiently in systems having a broad spectrum of phonon frequencies and many degrees of freedom.  We test and illustrate the new approach with an application to the phase XI of water ice using density-functional theory. 
We find that the SCHA reproduces extremely well the experimental thermal expansion of ice in the whole temperature range between $\SI{0}{\kelvin}$ and $\SI{270}{\kelvin}$, in contrast with the results obtained within the quasi-harmonic approximation, that underestimates the effect by about 25\%.
\end{abstract}

\maketitle
\section{\label{sec:intro}Introduction}
Atomic vibrations play a main role in many branches of physics and chemistry, as they are involved in
thermodynamic, transport, and superconducting proprieties of materials and molecules.
Many spectroscopic techniques, such as Raman and IR, measure how atoms vibrate. The standard approach to describe
vibrations is the harmonic approximation, in which the Born-Oppenheimer (BO) energy surface is approximated as a $3N$-dimensional
paraboloid around the ionic positions.
The solutions of the harmonic Hamiltonian are well-defined non-interacting vibrational quasiparticles, phonons,
with infinite lifetime and temperature independent spectrum.
Anharmonic effects, due to higher orders in the BO energy surface, introduce interaction between phonons.
As a result, phonons acquire a  finite lifetime that is responsible for thermal transport. Furthermore, phonon spectra become temperature dependent.

Anharmonic effects are commonly accounted by perturbation theory, whose validity range is limited only when the harmonic contribution dominates in the range defined by the quantum zero-point motion (ZPM).
This is not the case of many interesting phenomena, such as systems undergoing a displacive second order structural phase transition in 
which a phonon branch softens as a function of temperature, e.g. charge-density-waves (CDW) and ferroelectrics\cite{volkov1983,tsang1976,pawley1966diatomic,delaire2011giant,luspin1980soft,weber2011electron,leroux2015strong,holt2001x,leroux2012anharmonic,ghosez1998ab,calandra2009effect,calandra2011charge}, or 
in solids largely affected by the ZPM, for example in hydrides or in molecular crystals containing $\ch{H}$, like water and
high-pressure phases of hydrogen\cite{ErreaRev2016,stritzker1972superconductivity,ErreaB2013,Errea2014,Errea2015,schirber1974concentration}.
Classical molecular dynamics (MD) for ions or methods based on it can be used to extract the non-perturbative anharmonic renormalized phonon dispersion\cite{car1985,wang1990tight,Magdau_2013,ljungberg2013temperature,teweldeberhan2010high,zhang2014phonon,hellman2011lattice,hellman2013temperature,hellman2013temperature2}. However, within these approaches, quantum effects on nuclei are neglected. These methods are then inappropriate below the Debye temperature.

In order to correctly account for both quantum and anharmonic effects, the ideal technique is path-integral molecular dynamics (PIMD)\cite{Chandler_1981,Barker_1979,Ceriotti_2010}, but its demanding computational cost limits its applicability to systems with few atoms or to the use of empirical potentials. To overcome these problems many self-consistent approximations have been developed\cite{Tadano,MonserratSCF}. Among them, the self-consistent harmonic approximation (SCHA) allows one to describe anharmonicity through a full-quantum variational theory. The stochastic implementation of the SCHA~\cite{Errea2014} (SSCHA) allows us to apply the powerful variational SCHA method to many systems with a lower numerical effort than MD and PIMD, making it possible
the calculation of non-perturbative anharmonic effects from first-principles.

So far, the applications of the SSCHA method\cite{ErreaB2013,Errea2014, Errea2015,Borinaga2016,Nature2016,Bianco2017} have been limited to simple systems with high symmetry. The main reason is that the variational minimization as formulated in Ref.[\onlinecite{Errea2014}] can yield runaway solutions and become very inefficient in complex crystals that show a wide range of phonon frequencies and many degrees of freedom. Another limitation of the original SSCHA formulation is that it needs finite-difference approaches to estimate the effect of ionic fluctuations in the stress tensor, as it happens in the quasi-harmonic approximation, which is extremely cumbersome for non-cubic crystals. This hinders cell relaxation within the SSCHA.

In this work we efficiently overcome these difficulties by introducing a new equation for the stress tensor within the SCHA.
Furthermore, we develop a more robust minimization algorithm based on an analytical preconditioner combined with a nonlinear change of variables that allows for efficient many variables minimizations. Our developments pave the way to primitive cell relaxations including quantum and anharmonic effects avoiding finite difference approaches.
% Furthermore, the new minimization algorithm makes the SSCHA minimization possible in complex systems with many degrees of freedom, e.g. molecular systems, widening the applications of the method.

We illustrate and benchmark the method with the phase XI of ice (\ch{H2O}),
the perfect prototype of a complex molecular crystal. Ice XI is
 the ordered phase of common ice formed below 72~K in the presence of a small amount of an alkali hydroxide~\cite{Tajima1982}.
It is commonly used to study quantum effects in water thanks to its great similarity to normal ice (Ih)\cite{Pamuk_2012,Umemoto_2015}.
Ice is characterized by the interplay between intra-molecular covalent OH bonds and inter-molecular hydrogen bonds. The great difference in strength of inter-molecular and intra-molecular forces makes ice phases acquire a very broad spectra for their vibrational energies, from the very low energy rotons to the large energy vibrons. 
Moreover, this structure of ice experimentally exhibits at low temperature the negative thermal expansion\cite{Rottger1994} and the ``anomalous isotope volume effect''\cite{Pamuk_2012, Umemoto_2015,Rottger1994,Rottger2012}: if hydrogen is replaced by deuterium, the crystal volume expands in about a 0.1 \%. This is the opposite behaviour
of what is usually observed when  heavier isotope is substituted in the crystal.
These features make the XI phase of crystal ice a perfect benchmark for the here introduced SSCHA algorithm (\secname~\ref{sec:forcefield} and \ref{sec:abinitio}).

This work is organized as follows. We recall the basis of the SCHA algorithm in \secname~\ref{sec:SSCHA}. We introduce the stress tensor in the SCHA formalism in \secname~\ref{sec:stress}. We discuss about the stochastic implemetation of the algorithm in \secname~\ref{sec:stochastic}. 
We face the issues of the SSCHA minimization algorithm in \secname~\ref{sec:runaway}: we get an ansatz on the condition number of the minimization process (\secname~\ref{sec:hessian}), and provide two changes of variables that suppress it (\secname~\ref{sec:root} and~\ref{sec:precond}). Then, we benchmark the new SCHA algorithm in ice XI in \secname~\ref{sec:forcefield}. Finally, \secname~\ref{sec:abinitio} reports the results computed with density functional theory (DFT) in the unit cell of ice XI, compared with the quasi harmonic approximation (QHA). In \secname~\ref{sec:conclusions} we
summarize the main results of this work. The manuscript is completed with three appendices, where the mathematical derivations of the presented equations are provided.

\section{\label{sec:SSCHA}The self-consistent harmonic approximation}
The SCHA is a variational principle on the Bohr-Oppenheimer (BO) free energy.
The nuclear quantum Hamiltonian of a generic system can be defined in the BO approximation as
\beq
H = \sum_{n = 1}^N \sum_{\alpha = 1}^3 \frac{ {p_n^{\alpha}}^2}{2M_n} + V(\vec R, \left\{\vec a_i\right\}),
\eeq
where $V$ is the BO energy surface,  $M_n$ is the mass of the $n$-th atom, $p_n^\alpha$ and  $\vec R$ ($R_n^\alpha$) are the momentum and position operators of the nuclei in the periodic cell (or supercell), $N$ is the number of atoms,
and  $\left\{\vec a_i\right\}$ are
the 3 unit cell vectors. The $\alpha$ index identifies the Cartesian coordinate.
Fixed the temperature $T$ and the volume (i.e. the cell vectors $\left\{\vec a_i\right\}$), the free energy of the ionic Hamiltonian $H$ is:
\beq
F(\left\{\vec a_i\right\}) = \braket{H}_{\rho_H} + k_bT\braket{\ln \rho_H}_{\rho_H},
\label{eq:true:fe}
\eeq
where $\rho_H$ is the equilibrium density matrix
\beq
\rho_H = \frac{e^{-\beta H}}{\Tr e^{-\beta H}}\qquad \beta = \frac{1}{k_b T},
\eeq
and the brackets $\braket{ O}_{\rho_{H}}$ indicate the average of the observable $O$ according to the $\rho_H$ density matrix:
\beq
\braket{ O}_{\rho_{H}} = \Tr \left[\rho_H O\right].
\eeq
The equilibrium density matrix satisfies the free energy least principle. Given a trial density matrix $\rho_{\mathcal H}$, we can define a free energy functional 
whose minimum is the free energy:
\beq
\mathcal F(\left\{\vec a_i\right\})[\rho_{\mathcal H}] =  \braket{H}_{\rho_{\mathcal H}} + k_b T\braket{\ln \rho_{\mathcal H}}_{\rho_{\mathcal H}},
\eeq
\beq
F(\left\{\vec a_i\right\}) = \min_{\rho_{\mathcal H}}\mathcal  F(\left\{\vec a_i\right\})[\rho_{\mathcal H}].
\eeq
The SCHA approximation consists in the restriction of the possible trial density matrices to the equilibrium one obtained from a harmonic Hamiltonian:
\begin{subequations}
\beq
\mathcal H_{\vec {\mathcal R}, \mathbf{ \Phi}} =  \sum_{n = 1}^N \sum_{\alpha = 1}^3 \frac{ {p_n^{\alpha}}^2}{2M_n} + \mathcal V_{\mathbf{\Phi}, \vec {\mathcal R}}(\vec R),\mbox{ where}
\eeq
\beq
\mathcal V_{\mathbf{\Phi}, \vec {\mathcal R}}(\vec R) = \frac 1 2 \sum_{\substack{n = 1\\ m= 1}}^N \sum_{\substack{\alpha = 1\\ \beta = 1}}^{3}
u_n^{\alpha} \Phi_{nm}^{\alpha\beta}  u_m^{\beta},\mbox{ and}
\eeq
\beq
u_n^{\alpha} = R_n^\alpha - {\mathcal R}_n^{\alpha}.
\eeq
\beq
\rho_{\mathcal H} = \rho_{\vec {\mathcal R}, \mathbf{\Phi}} = \frac{e^{-\beta \mathcal H_{\vec {\mathcal R}, \mathbf{\Phi}}}}{\Tr e^{-\beta\mathcal H_{\vec {\mathcal R}, \mathbf{\Phi}}}}.
\eeq
\end{subequations}
Here $u_{a}^\alpha$ is the displacement of $a$-th atom along the $\alpha$ direction respect to a central position $\vec {\mathcal R}$, and
$\Phi^{\alpha\beta}_{nm}$ is the matrix element of the real space force constant matrix (we use the bold font to indicate tensors and matrices). 
With the introduction of the auxiliary harmonic Hamiltonian it is possible to recast the free energy as
\beq
\fscha= F_{\mathbf \Phi} + \braket{V - \mathcal V_{\mathbf \Phi, \vec{\mathcal R}}}_{\rho_{\vec {\mathcal R},\mathbf{\Phi}}},
\label{eq:f:func:scha}
\eeq
where $F_{\mathbf \Phi}$ is the exact free energy of the harmonic Hamiltonian:
\beq
\label{eq:func:f}
F_{\mathbf \Phi}(T) = \sum_{\mu = 1}^{3N} \left[\frac{\hbar\omega_\mu}{2} + \frac 1 \beta \ln\left(1 - e^{-\beta \hbar\omega_\mu}\right)\right],
\eeq
where $\omega_\mu$ and $\vec e_\mu$ are, respectively, the eigenvalues and eigenvectors of the $\mathbf \Phi$ matrix divided by the atomic masses:
\beq
\sum_{t = 1}^N \sum_{\beta = 1}^3 \frac{\Phi^{\alpha\beta}_{st}}{\sqrt{M_sM_t}} {e_{\mu}}_t^\beta = \omega_{\mu}^2 {e_\mu}_s^\alpha.
\label{eq:scha:freq}
\eeq

The real free energy can, therefore, be approximated as the minimum of the free energy functional (\eqname~\ref{eq:f:func:scha}) with respect to $\vec {\mathcal R}$ 
and $\mathbf\Phi$: 
\begin{subequations}
\beq
\mathcal F({\vec {\mathcal R}}, \left\{\vec a_i\right\}) = \min_{\mathbf \Phi} \fscha,
\label{eq:scha:semimin}
\eeq
\beq
\mathcal F(\left\{\vec a_i\right\}) = \min_{\mathbf \Phi, \vec{\mathcal R}}  \fscha.
\label{eq:scha:min}
\eeq
\end{subequations}
From now on, when we drop one of $\vec {\mathcal R}$ or $\mathbf\Phi$ symbols, we mean the quantity computed in the value of that variable that minimizes the free energy. For example, the equilibrium SCHA density matrix is just $\rho$.

One of the advantages of using the harmonic Hamiltonian to restrict the $\rho_{\mathcal H}$ space is that we have a trivial physical interpretation of the minimization parameters. In fact
$\vec {\mathcal R}$ represents the centroid positions of the atoms, i.e. the anharmonic average positions as measured by, e.g., diffraction experiments:
\beq
\vec {\mathcal R}  = \braket{\vec R}_\rho.
\eeq
In the same way, $\mathbf{\Phi}$ is related to the thermal and quantum fluctuations and defines the real space density matrix broadening. Within the harmonic auxiliary Hamiltonian, the probability distribution function defined by the real space density matrix is a product of Gaussians:
\begin{subequations}
\beq
\rho_{\vec {\mathcal R}, \mathbf{\Phi}}(\vec u) = \braket{\vec u | \rho_{\vec {\mathcal R} \mathbf{\Phi}} | \vec u}\nonumber
\eeq
\beq
\rho_{\vec {\mathcal R}, \mathbf{\Phi}}(\vec u) = \sqrt{\det\left( \mathbf{\Upsilon} /2\pi\right)} \exp\left( - \frac 12 \sum_{st\alpha\beta} \Upsilon_{st}^{\alpha\beta} u_s^\alpha u_t^\beta\right)
\label{eq:rho}
\eeq
where
\beq
\Upsilon_{st}^{\alpha\beta} = \sqrt{M_s M_t} \sum_\mu \frac{2\omega_\mu}{(1 + 2n_\mu) \hbar} {e_\mu}_s^\alpha {e_\mu}_t^\beta
\label{eq:upsilon}
\eeq
\end{subequations}
and $n_\mu$ are the boson average occupation number for the $\mu$ mode. It is important to notice that $\omega_\mu$ and $\vec e_\mu$ (\eqname~\ref{eq:scha:freq}) 
are not directly equal to the physical phonons since they are constrained to be positive defined\cite{Bianco2017}. Instead, they are related to quantum and thermal fluctuations: they uniquely define the $\bm\Upsilon$ tensor. 

It is possible to define the SCHA force as the derivative of the
free energy (\eqname~\ref{eq:scha:semimin}) with respect to the nuclear average positions,
\beq
\label{eq:scha:forces}
- \frac{\partial \mathcal F}{\partial {\mathcal R}_n^\alpha}(\vec {\mathcal R}, \left\{\vec a_i\right\}) = \braket{f_n^\alpha - {f_{\mathcal H}}_n^\alpha}_{\rho_{\vec{\mathcal R}}},
\eeq
where $\vec f$ and $\vec {f_{\mathcal H}}$ are, respectively, the BO and harmonic forces:
\beq
f_n^\alpha = - \frac{\partial V}{\partial R_n^\alpha}(\vec R, \left\{\vec a_i\right\})
\eeq
\beq
{f_{\mathcal H}}_n^\alpha = - \frac{\partial\mathcal V_{\vec{\mathcal R}, \mathbf\Phi}}{\partial R_n^\alpha} = - \sum_{m = 1}^{N} \sum_{\beta = 1}^3 \Phi_{nm}^{\alpha\beta} u_m^\beta
\eeq
It is interesting to notice how the harmonic potential $\mathcal V_{\vec {\mathcal R}, \mathbf\Phi}$ does not depend explicitly on the unit cell vectors $\left\{\vec a_i\right\}$, while the BO energy $V(\vec R, \left\{\vec a_i\right\})$ does.

%% THE SD AND CG MINIMIZATION
To numerically minimize the SCHA free energy it is possible to use the steepest descend (SD) or conjugate gradient (CG) methods\cite{press2007numerical}, both based on the knowledge of the
gradient of the function to minimize. This can be expressed as a function of the averages of the BO and harmonic forces\cite{Errea2014}
\begin{subequations}
\label{eq:gradients}
\begin{align}
  \nonumber \nabla_{\bm \Phi} \fscha & = - \sum_{st \alpha\beta\mu}
  \sqrt\frac{M_t}{M_s} \left({e_\mu}_s^{\alpha}\nabla_{\bm \Phi} \ln a_\mu + \nabla_{\bm \Phi} {e_\mu}_s^\alpha\right)
  \\
  & \times \braket{ \left[f_s^\alpha(\vec u) - {f_{\mathcal H}}_s^\alpha(\vec u)\right] u_t^\beta}_\rscha {e_\mu}_t^\beta,
\label{eq:gradient}
\end{align}
\beq
\nabla_{{\mathcal R}_s^\alpha}\fscha = - \braket {f_s^\alpha - {f_{\mathcal H}}_s^\alpha}_\rscha,
\label{eq:wy:g}
\eeq
\end{subequations} 
% and from the second derivative of the free energy it is possible to obtain the SCHA self-consistent relation\cite{Bianco2017}:
% \beq
% \frac{\partial^2 \fscham}{\partial {\mathcal R}_s^\alpha \partial {\mathcal R}_t^\beta} = \braket{ \frac{\partial^2 V(\vec R, \left\{\vec a_i\right\})}{\partial R_s^\alpha \partial R_t^\beta}}_\rscham= \Phi_{st}^{\alpha\beta}.
% \label{eq:self:consistent}
% \eeq

In the next section we show how to implement SCHA in an isobaric ensemble, allowing for the relaxation also of the unit cell. This is achieved thanks to the introduction of the stress tensor in the SCHA framework.

\section{The stress tensor in the self-consistent stochastic approximation}
\label{sec:stress}
To minimize the free energy with respect to the lattice parameters in a periodic system, the knowledge of the stress tensor is crucial. 
The SCHA stress can be defined as:
\beq
\label{eq:p0}
P_{\alpha\beta}(\vec {\mathcal R}, \left\{\vec a_i\right\}) = - \frac{1}{\Omega}\left. \frac{\partial \mathcal F(\vec {\mathcal R}, \left\{\vec a_i\right\})}{\partial\varepsilon_{\alpha\beta}} \right|_{\bm {\varepsilon} = 0},
\eeq
where $\Omega$ is the volume of the system and the strain tensor $\varepsilon_{\alpha\beta}$ identifies a generic deformation, where both the lattice parameters and the average central position are affected:
\begin{subequations}
\label{eq:strain:def}
\beq
 {a_i'}^\alpha = {a_i}^\alpha +  \sum_{\beta = 1}^3 \varepsilon_{\alpha\beta} {a_i}^\beta,
\eeq
\beq
{\mathcal {R}'}^\alpha_n = {\mathcal R}_n^\alpha + \sum_{\beta = 1}^3 \varepsilon_{\alpha\beta} {\mathcal R}_n^\beta.
\eeq
\end{subequations}
% The derivative~\eqref{eq:p0} can be rewritten in the variables $\mathcal F_{H}$ explicitly depends on (\eqname~\ref{eq:func}):
% \beq
% P_{\alpha\beta} = - \frac{1}{\Omega} \left[\sum_{s\gamma} \frac{\partial\mathcal F_{H}}{\partial{R_c}^\gamma_s} \frac{\partial {R_c}^\gamma_s}{\partial\varepsilon_{\alpha\beta}} + \sum_{i\gamma}\frac{\partial F_{H}}{\partial {a_i}^\gamma} \frac{\partial {a_i}^\gamma}{\partial \varepsilon_{\alpha\beta}}\right]
% \eeq
This is equivalent to performing a strain keeping fixed the internal crystal coordinates of the system. The final result can be divided into three main contributions
(see Appendix~\ref{app:stress1} for the proof):
\begin{eqnarray}
\label{eq:p:final}
P_{\alpha\beta}(\vec{\mathcal R},\left\{\vec a_i\right\}) = P_{\alpha\beta}^H(\vec{\mathcal R}, \left\{\vec a_i\right\}) + P_{\alpha\beta}^{\mbox{\scriptsize FLC}}(\vec{\mathcal R}, \left\{\vec a_i\right\}) + \nonumber\\P_{\alpha\beta}^{\mbox{\scriptsize FRC}}(\vec{\mathcal R}, \left\{\vec a_i\right\}) ,
\end{eqnarray}
where the $P_{\alpha\beta}^H(\vec{\mathcal R})$ is the static contribution, i.e. the stress tensor computed without quantum and thermal fluctuations (classical with $T=0$), $P_{\alpha\beta}^{\mbox{\scriptsize FLC}}$ is the contribution of the fluctuations to the stress, and  $P_{\alpha\beta}^{\mbox{\scriptsize FRC}}$ is an extra term that takes into account the work necessary to move the centroids according to the applied strain $\bm \varepsilon$:
\begin{widetext}
\begin{subequations}
\label{eq:stress:pieces}
\beq
\label{eq:classical:stress}
P^{H}_{\alpha\beta}(\vec R, \left\{\vec a_i\right\}) = -\frac{1}{\Omega}\left. \frac{\partial V(\vec R, \left\{\vec a_i\right\})}{\partial\varepsilon_{\alpha\beta}}\right|_{\bm \varepsilon = 0},
\eeq
\beq
 P_{\alpha\beta}^{\mbox{\scriptsize FLC}}(\vec{\mathcal R}, \left\{\vec a_i\right\}) = \braket{P_{\alpha\beta}^H(\vec R, \left\{\vec a_i\right\})}_{\rscham} - P_{\alpha\beta}^H(\vec {\mathcal R}, \left\{\vec a_i\right\}) - \frac{1}{2\Omega} \sum_{s = 1}^N\braket{ \left({f_{\mathcal H}}_s^\alpha u_s^\beta + {f_{\mathcal H}}_s^\beta u_s^\alpha\right)}_\rscham,
\label{eq:p:fluct}
\eeq
\beq
 P_{\alpha\beta}^{\mbox{\scriptsize FRC}}(\vec{\mathcal R}, \left\{\vec a_i\right\}) = 
\frac{1}{2\Omega} \sum_{s = 1}^N \left({\mathcal R}_s^\beta \braket{f_s^\alpha - {f_{\mathcal H}}_s^\alpha}_\rscham + {\mathcal R}_s^\alpha \braket{f_s^\beta - {f_{\mathcal H}}_s^\beta}_\rscham\right).
\label{eq:p:force}
\eeq
\end{subequations}
\end{widetext}
The last term in \eqname~\eqref{eq:p:fluct} makes fluctuations on pressure disappear in a pure harmonic crystal (see appendix~\ref{app:stress1}). The force term, i.e. \eqname~\eqref{eq:p:force}, is non-zero only if the SCHA centroids $\vec{\mathcal R}$ are not in the equilibrium configuration (i.e. the SCHA force \eqname~\ref{eq:scha:forces} on each atom is not zero), and it is independent of the choice of the origin (the sum of the forces over atom indexes is zero).

\eqname~\eqref{eq:stress:pieces} can be computed once we know the BO surface  $V(\vec R, \left\{\vec a_i\right\})$, the atomic force $f_s^\alpha(\vec R, \left\{\vec a_i\right\})$ and the stress tensors $P^H_{\alpha\beta}(\vec R,\left\{\vec a_i\right\})$ for each ionic displacement $\vec R$ in the ensemble of the configurations distributed according to $\rscham(\vec R)$. In \secname~\ref{sec:stochastic} we discuss an efficient stochastic implementation to numerically compute this average. 
The computation of the SCHA stress tensor enables the complete unit cell relaxation in isobaric conditions (fixing the external $P^*$ pressure). This is done by minimizing
the Gibbs free energy, that is obtained from the Helmholtz free energy through the Legendre transform
\beq
\mathcal G(\vec{\mathcal R},P^*) = \fscham + P^*\,\Omega(\left\{\vec a_i\right\}).
\eeq

\section{The stochastic implementation}
\label{sec:stochastic}
The SCHA algorithm can be implemented by performing the stochastic evaluation of all the averages. Thanks to the fact that the density matrix is a multidimensional Gaussian function (\eqname~\ref{eq:rho}), it is possible to generate an ensemble distributed according to $\rho_{\vec {\mathcal R},\mathbf{\Phi}}$ without any Metropolis algorithm\cite{Errea2014}, and the average of a generic observable $O(\vec R)$ can be computed through Monte Carlo integration:
\beq
\braket{O(\vec R)}_{\rho_{\vec {\mathcal R}, \mathbf{\Phi}}} = \frac{1}{N_c} \sum_{\vec R_I} O(\vec R_I),
\eeq
where $N_c$ and $\vec R_I$ are, respectively, the dimension  and the configurations of the ensemble. To avoid regenerating the ensemble at each minimization step
it is convenient to introduce the importance sampling reweighting\cite{Errea2014}:
\beq
\rho_I =  \frac{\rho_{\vec {\mathcal R}, \mathbf{\Phi}}(\vec R_I)}{\rho_{\vec {\mathcal R}_{SG}, \mathbf{\Phi}_{SG}}(\vec R_I)}
\eeq
where $\rho_{\vec {\mathcal R}_{SG}, \mathbf{\Phi}_{SG}}(\vec R_I)$ is the density matrix used to extract the ensemble configurations, i.e. computed with the starting guess for the centroid positions $\vec{\mathcal R}_{SG}$ and the auxiliary dynamical matrix $\mathbf \Phi_{SG}$. Then, the average of the observable
$O$ in a generic value of $\vec{\mathcal R}$ and $\mathbf \Phi$ can be computed through:
\beq
\label{eq:is:average}
\braket{O(\vec R)}_{\rho_{\vec {\mathcal R}, \mathbf{\Phi}}} = \frac{1}{N_c} \sum_{\vec R_I} \rho_I O(\vec R_I).
\eeq

The reweighting procedure allows us to overtake the usually high computational effort required by the SSCHA minimization. In fact, the computation of the SCHA
free energy gradient\cite{Errea2014}, as well as the SCHA stress tensor (\eqname~\ref{eq:stress:pieces}), requires only the knowledge of first derivative of
the BO energy in the ensemble, that can be obtained just in one total energy calculation per configuration thanks to the Hellmann-Feynman theorem.
Moreover, the total energy calculation can be computed only one time in the starting ensemble of configurations $\vec R_I$, and then recycled on the whole
minimization thanks to the reweighting. When the new variables $\vec{\mathcal R}$ and $\bm\Phi$ are too distant from the initial ones, $\vec{\mathcal R_{SG}}$
and $\bm\Phi_{SG}$, the ensemble is no more able to provide a good estimation of the stochastic averages and it must be re-extracted.
Thus, the overall computational 
effort to run an ``ab-initio'' SSCHA calculation is given by the number of times the initial ensemble is regenerated. 

It is crucial to improve the reliability of the ensemble, in order to minimize the number of times the initial ensemble is regenerated during the SSCHA free energy optimization. To this purpose, we introduce both a \emph{symmetrized sampling} and a new stochastic threshold to evaluate the importance sampling accuracy. 

The real space density matrix is a symmetric distribution, $\rscha(\vec u) = \rscha(-\vec u)$, and all observables required in the SSCHA free energy minimization
are purely even or odd terms of
the Taylor expansion of  $V(\vec R, \left\{\vec a_i\right\})$ in $(\vec R - \vec{\mathcal R})$. 
To reduce the stochastic noise we implemented the \emph{symmetrized sampling}\cite{Rios2017}: for each displacement $\vec u$ generated, also its opposite $-\vec u$ is included in the ensemble. This analytically cancels all the non contributing terms in the Taylor expansion of the BO energy.
It is important to notice that this advantage is lost when $\vec{\mathcal R} \neq \vec{\mathcal R}_{SG}$. However, we still find the \emph{symmetrized sampling} to
be convenient to reduce the stochastic noise even if the centroids do not match perfectly the starting guess.

The previous estimator of the importance sampling accuracy used by Errea \emph{et al.}~\cite{Errea2014} was the check on the $\rho_I$ normalization:
\beq
\left | \frac {1}{N_c} \sum_{I = 1}^{N_c} \rho_I - 1 \right| < \eta.
\label{eq:eta1}
\eeq
However, this threshold can be exceeded if all the weight constantly drift from the uniform value, or it can remain satisfied if they spread a lot. Thus, a 
much better estimator that consider the spreading of the different configuration weights can be implemented.
In order to improve the reliability of the reweighting procedure we found more reliable the Kong-Liu effective sample size~\cite{Kong1994}:
\beq
\label{eq:kl}
N_{eff} = \frac{\left(\sum_I \rho_I\right)^2}{\sum_I \rho_I^2}< N_c.
\eeq
A new critical threshold $\eta'$ can be defined as:
\beq
\label{eq:eta2}
\frac{N_eff}{N_c} > \eta'.
\eeq
If the weight $\rho_I$ of a configuration goes to zero, it does not contribute to the averages.
The effective sample size counts how many configurations are actually contributing to the Monte Carlo average (\eqname~\ref{eq:is:average}), even if the $\rho_I$ are
properly normalized.

In all the simulations reported in this work we set $\eta' = 0.6$. If the critical threshold is overcome, the minimization is stopped and a new ensemble is generated with the final trial density matrix $\rho_{\vec {\mathcal R}, \mathbf{\Phi}}$.

\section{Minimization instabilities and runaway solutions}\label{sec:runaway}
The SSCHA algorithm consists in minimizing the free energy through the stochastic evaluation of its gradient (\eqname~\ref{eq:gradients}), employing SD or CG algorithm,
and taking advantage of  reweighting to perform multiple SD or CG steps without recomputing energies and forces of the ensemble at each step.
However, this minimization procedure  was empirically found to be very difficult in some systems, especially in those near a structural instability, where a phonon mode frequency softens to zero, while all the other modes are substantially higher in energy, or molecular crystals, in which hard intermolecular vibrations coexist with low energy intramolecular modes. In these cases, of very great physical interest, the stochastic free energy minimization requires a large number of ensemble regeneration (and consequently a large number of
first principles force calculations) to converge. Moreover, the minimization can lead to ``runaway solutions'': fake nonphysical solutions of the SCHA self-consistency where the auxiliary dynamical matrix $\mathbf \Phi$ is not positive defined. 

To understand the convergence properties we consider the SCHA free energy close to the minimum. It can be approximated as a quadratic form in the minimization variables ($\vec{\mathcal R}$ and $\mathbf\Phi$). Under this condition, the SCHA free energy is expressed by the Hessian matrix $\bm A$ with respect those variables. From the Hessian matrix it is possible to define the condition number\cite{MauriPrecond} $C$, as:
\beq
\label{eq:condition:number}
C = \frac{\max \lambda_{\bm A}}{\min\lambda_{\bm A}},
\eeq
where $\lambda_{\bm A}$ is a generic eigenvalue of the $\bm A$ matrix. In the limit in which the number of degrees of freedom is much greater than the number of minimization steps, the steepest descent (SD) and the conjugate gradient (CG) algorithms converge into a fixed threshold with almost $N$ steps proportional\cite{press2007numerical} to:
\begin{subequations}
\beq
N_{SD} \propto C
\eeq
\beq
N_{CG} \propto\sqrt C
\eeq
\end{subequations}

In the SCHA case, the number of minimization steps is proportional to the number of times the critical threshold $\eta'$ is overcome. Then, this number must be carefully optimized, since each time the ensemble is re-extracted, the ab-initio energies and forces for each configuration must be computed. This
calculation is the overall computational cost of the algorithm.
In the next sections we provide an ansatz for the condition number, unveiling that it dramatically diverges in the aforesaid cases. 
We further provide two ways to prevent this divergence, paving the way to the application of SSCHA in these systems.

\subsection{Hessian matrix}\label{sec:hessian}
In this section we provide an analytical guess of the free energy Hessian matrix $\bm A$ with respect to the minimization variable $\bm\Phi$. In general, this is not possible, since
computing the real Hessian matrix corresponds to solving exactly the problem. However, we can perform the computation in an analytical test case that, hopefully, will
enclose all the physics of the minimization problems incurred so far. This is a purely harmonic system, described by an harmonic Hamiltonian.
From now on we introduce a compact notation to describe both Cartesian and atomic indices ($v_a = v_s^\alpha$). 
\beq
H = \frac 12 \sum_{a}\frac{\left(p_a\right)^2}{2M_{a}} + \frac 1 2\sum_{ab} u_a K_{ab} u_b,
\eeq

The free energy Hessian matrix with respect to the $\mathbf \Phi$ variable can be computed analytically. The steps that lead to the following result are reported in appendix~\ref{app:hessian}:
\beq
\label{eq:hessian}
A_{\bm \Phi}^{abcd} = \left.\frac{\partial^2 \fscha}{\partial\Phi_{ab}\partial\Phi_{cd}}\right|_{ \mathbf \Phi = \mathbf K} = \frac 1 2 \mathcal P_{ab}\mathcal P_{cd}\left(\Lambda^{abcd} + \Lambda^{abdc}\right),
\eeq
where the $\bm\Lambda$ 4-rank tensor is the same introduced by Bianco \emph{et al.}~\cite{Bianco2017}, and $\mathcal P$ is a symmetrization factor
\begin{align}
\nonumber  \Lambda^{abcd}& = -\frac{\hbar}{4}\sum_{\mu\nu}\frac{1}{\omega_\mu\omega_\nu}\frac{e_\nu^a e_\mu^b e_\nu^c e_\mu^d}{\sqrt{M_aM_bM_cM_d}}\times \\
  & \times\left\{\begin{array}{ll} \displaystyle \frac{2n_\nu + 1}{2\omega_\nu} - \frac{dn_\nu}{d\omega_\mu} & \omega_\nu = \omega_\mu \\ \\ \displaystyle
  \frac{n_\mu + n_\nu + 1}{\omega_\mu + \omega_\nu} - \frac{n_\mu - n_\nu}{\omega_\mu - \omega_\nu} & \omega_\nu \neq\omega_\mu\end{array}\right.,
\end{align}
\beq
\mathcal P_{ab} = \sqrt 2 \left(1 - \delta_{ab}\right) + \delta_{ab}.
\eeq
Here the $\omega_\mu$ and $\vec e_\mu$ are the frequencies and polarization vectors of the $\bm K$ matrix. These are, indeed, equal to the $\bm\Phi$ matrix
in the minimum of the SCHA free energy, and represent the real phonons of the system.

The $\bm \Lambda$ matrix can be diagonalized analytically if we consider the case of all equal masses
\beq
\sum_{cd}\Lambda^{abcd}e_\mu^c e_\nu^d = \tilde \lambda_{\mu\nu} e_\mu^a e_\mu^b.
\eeq
We can obtain an easy expression of the spectrum of the Hessian matrix in the pure quantum limit $T\rightarrow 0$ and the pure classical
limit $T \rightarrow \infty$:
\begin{subequations}
\beq
\lim_{T\rightarrow 0}\tilde \lambda_{\mu\nu} = - \frac {\hbar}{ 4 M^2} \frac{1}{\omega_\mu\omega_\nu(\omega_\mu + \omega_\nu)},
\eeq
\beq
\lim_{T \rightarrow \infty} \tilde\lambda_{\mu\nu} =
-\frac{1}{4\beta M^2}\frac{1}{\omega_\mu^2\omega_\nu^2}\left[1 + \frac{\omega_\mu\omega_\nu}{(\omega_\mu + \omega_\nu)^2}\right]. 
\eeq
\end{subequations}
Therefore, the Hessiam matrix spectrum goes as  $\omega^{-3}_\mu$ in the quantum limit and $\omega^{-4}_\nu$ in the classical one.
We can compute the
 condition numbers, as defined in \eqname~\eqref{eq:condition:number}:
\begin{subequations}
\label{cond:numbers}
\beq
C_{\bm\Phi, T = 0} \approx \left(\frac{\omega_{max}}{\omega_{min}}\right)^3 ,
\eeq
\beq
C_{\bm\Phi,T\rightarrow\infty} \approx \left(\frac{\omega_{max}}{\omega_{min}}\right)^4.
\eeq
\end{subequations}
This unveils the pathology in the SSCHA minimization if the gradient is taken with respect to $\bm\Phi$ as presented in Ref.[\onlinecite{Errea2014}] for the mentioned systems: 
when we have a structural instability, there is a phonon mode that softens to zero ($\omega_{min} \rightarrow 0$),  producing a diverging
condition number $C\rightarrow\infty$. 
In the same way, molecular crystals have a broad spectrum, with very large difference between the highest vibron modes and the lowest intermolecular ones (for example in common ice we have $\omega_{max}/\omega_{min} \sim 10^{3}$). 
This yields  extremely high values of the condition numbers, that makes the minimization really difficult, and
requires lots of energy and force recalculations to achieve a good minimization. This obviously hinders the application of the SSCHA method fully
``ab-inito'' in complex systems.

\subsection{Nonlinear change of variable}\label{sec:root}
The condition number is a function of the minimization variables. Therefore, a simple change of variables can result in a powerful improvement in the
minimization algorithm. In this section we show that it is possible to almost completely solve the divergences occurring in the condition numbers (\eqname~\ref{cond:numbers}) with a simple nonlinear change in the $\bm \Phi$ auxiliary matrix.
Moreover, we can also completely cancel the aforesaid ``runaway solutions''.
The ``runaway solutions'' are fake non-physical solutions of the SCHA that may arise during the minimization if the $\bm\Phi$ matrix is not positive definite.
In order to avoid this problem one could perform a constrained minimization. It is difficult to implement this kind of constrains  with SD or CG
algorithm. We find much more convenient to introduce a nonlinear change of variables, where we replace the auxiliary dynamical matrix $\mathbf\Phi$ with one of its even root:
\beq
\bm \Phi \rightarrow\sqrt[2n]{\mathbf\Phi}.
\eeq 
This mathematically constrains the minimization to have only positive defined matrices $\bm\Phi$. 
Does this nonlinear change improve the condition number on the minimization?

We can compute the Hessian matrix $\bm A_{\sqrt{\bm\Phi}}$ with respect to the square root of $\bm\Phi$ where $\bm\Phi$ minimizes the free energy:
\beq
\bm A_{\sqrt\mathbf \Phi} = \frac{\partial^2 \fscha}{\partial\sqrt\mathbf\Phi\partial\sqrt\mathbf\Phi} = \mathbf\Phi \bm A_{\mathbf \Phi} + 2 \sqrt\mathbf\Phi\bm A_{\mathbf\Phi} \sqrt\mathbf\Phi +\bm A_{\mathbf\Phi} \mathbf\Phi,
\label{eq:hessian:root}
\eeq
where $\bm A_{\mathbf\Phi}$ is the 4-rank Hessian with respect $\bm \Phi$ (\eqname~\ref{eq:hessian}). The procedure can be iterated to obtain any even root of $\bm\Phi$. 
Here we report also the $\sqrt[4]{\bm\Phi}$ expression, since, as we will show, it has a very favorable condition number:
\beq
\bm A_{\sqrt[4]{\mathbf \Phi}} = \sqrt\mathbf\Phi \bm A_{\sqrt\mathbf\Phi} + 2 \sqrt[4]{\mathbf\Phi}\bm A_{\sqrt\mathbf\Phi}\sqrt[4]{\mathbf\Phi} + \bm A_{\sqrt\mathbf\Phi}\sqrt\mathbf\Phi.
\label{eq:hessian:root4}
\eeq
We can easily compute the condition numbers in the new variables if all the masses are equal substituting~\eqname~\eqref{eq:hessian} 
into \eqref{eq:hessian:root} and \eqref{eq:hessian:root4} (recalling that $\bm\Phi \sim \omega^2$):
% \begin{subequations}
% \beq
% \label{eq:phi}
% \Phi_{ab} = \sum_\mu M \omega_\mu^2 e_{\mu}^a e_\mu^b,
% \eeq
% \beq
% \label{eq:sphi}
% \sqrt{\Phi}_{ab} = \sum_\mu \sqrt M \omega_\mu e_\mu^a e_\mu^b.
% \eeq
% \end{subequations}
\beq
C_{\sqrt[2]{\bm\Phi}, T=0} \sim \left(\frac{\omega_{max}}{\omega_{min}}\right)\qquad C_{\sqrt[2]{\bm\Phi}, T\rightarrow\infty} \sim \left(\frac{\omega_{max}}{\omega_{min}}\right)^2,
\eeq
\beq
C_{\sqrt[4]{\bm\Phi}, T=0} \sim 1\qquad C_{\sqrt[4]{\bm\Phi}, T\rightarrow\infty} \sim \left(\frac{\omega_{max}}{\omega_{min}}\right).
\eeq

The nonlinear change of variable $\bm\Phi\rightarrow\sqrt[4]{\bm\Phi}$ both avoids the non physical runaway solutions constraining the 
minimization space to admit only positive defined matrices, and strongly suppress the condition number making it independent on the phonon frequencies in
the $T = 0$ case, and suppressing it by a 4-th root in the classical case.

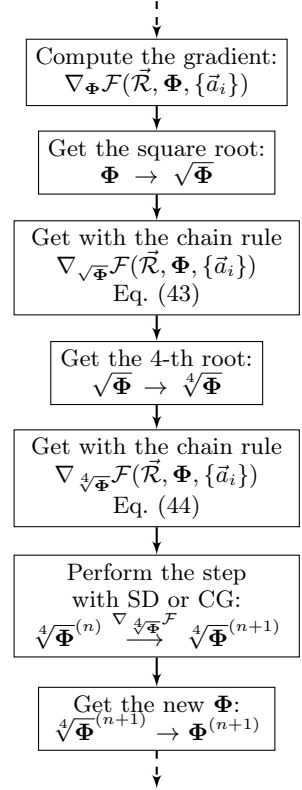
\begin{figure}
\centering
\begin{tikzpicture}[auto, block_center/.style={rectangle, draw=black, fill=white, text centered}, 
  line/.style={draw, thick, -latex', shorten >=0pt}]
\node [block_center, text width=10em] (Gradient) {Compute the gradient:\\ $\nabla_{\mathbf\Phi} \fscha$} ;
\node [block_center, text width=9em, node distance=1.2cm, below of=Gradient] (GetSqrt) {Get the square root:\\ $\bm\Phi \rightarrow \sqrt{\bm\Phi}$};
\node [block_center, text width=11em, node distance=1.4cm, below of=GetSqrt] (GetGradientSqrt) {Get with the chain rule \\$\nabla_{\sqrt{\bm\Phi}}\fscha$ \\\eqname~\eqref{eq:grad:sphi}};
\node [block_center, text width=8em, node distance=1.4cm, below of=GetGradientSqrt] (GetRoot4) {Get the 4-th root:\\ $\sqrt{\bm\Phi} \rightarrow \sqrt[4]{\bm\Phi}$};
\node [block_center, text width=11em, node distance=1.4cm, below of=GetRoot4] (GetGradientRoot4) {Get with the chain rule\\ $\nabla_{\sqrt[4]{\bm\Phi}}\fscha$ \\\eqname~\eqref{eq:grad:r4phi}};
\node [block_center, text width=11em, node distance=1.7cm, below of=GetGradientRoot4] (MinStep) {Perform the step with SD or CG:\\ $\sqrt[4]{\bm\Phi}^{(n)} \stackrel{\nabla_{\sqrt[4]{\bm\Phi}}\mathcal F}{\longrightarrow} 
\sqrt[4]{\bm\Phi}^{(n+1)}$};
\node [block_center, text width=9em, node distance=1.5cm, below of=MinStep] (GetPhi) {Get the new $\bm\Phi$:\\ $\sqrt[4]{\bm\Phi}^{(n+1)} \rightarrow \bm\Phi^{(n+1)}$};
\node [right= 1cm of GetRoot4] (most_east) {}; 
\begin{scope}%[every path./style=line]
\path[line, dashed]  ($(Gradient.north)+(0,0.5)$) -- (Gradient.north) ;
\path[line] (Gradient) -- (GetSqrt);
\path[line] (GetSqrt) -- (GetGradientSqrt);
\path[line] (GetGradientSqrt) -- (GetRoot4);
\path[line] (GetRoot4) -- (GetGradientRoot4);
\path[line]  (GetGradientRoot4) -- (MinStep);
\path[line] (MinStep) -- (GetPhi);
%\path[line] (GetPhi.east) -- ($(GetPhi.east)+(1,0)$) -- ($(Gradient.east)+(1,1)$) -- ($(Gradient)+(0,1)$) -- (Gradient.north);
\path[line,dashed] (GetPhi.south) -- ($(GetPhi.south)+(0,-0.5)$);
\end{scope}
\end{tikzpicture}
\caption{\small Flowchart on a minimization step with the $\mathbf{\Phi}\rightarrow \sqrt[4]{\bm\Phi}$ change of variables.\label{fig:flowchart:nonlinear}}
\end{figure}

In practice, the minimization in the $\sqrt[4]{\bm\Phi}$ is performed by computing the free energy gradient with respect to the new variable adopting the chain rule
on the derivatives:
\begin{align}
\nabla_{\sqrt\mathbf \Phi}\fscha &= \sqrt\mathbf\Phi\nabla_{\mathbf \Phi} \fscha\nonumber +\\
& +  \nabla_{\mathbf \Phi} \fscha \sqrt\mathbf\Phi,\label{eq:grad:sphi}
\end{align}
\begin{align}
\nabla_{\sqrt[4]{\mathbf\Phi}}\fscha &= \sqrt[4]{\mathbf \Phi}\nabla_{\sqrt\mathbf\Phi} \fscha\nonumber +\\
&+  \nabla_{\sqrt\mathbf\Phi}\fscha \sqrt[4]{\mathbf\Phi}.\label{eq:grad:r4phi}
\end{align}
The minimization step is updated as described by the flowchart reported in \figurename~\ref{fig:flowchart:nonlinear}

\subsection{Preconditioning}\label{sec:precond}
Even if the fourth root change of variable considerably improves the condition number, for high temperature calculations it still depends on the phonon frequencies linearly,
which could be problematic when a phonon mode goes close to zero near a structural phase transition. 
The SSCHA minimization algorithm correspond to finding the zeros of the free energy gradient:
\beq
\nabla_{\bm\Phi}\fscha = 0.
\eeq
From the above system, the SD and the CG algorithms are derived. However, 
since $\bm A_{\bm\Phi}$ is a positive defined matrix, the solution of the SCHA equation coincides with the solution of the auxiliary problem:
\beq
\bm A^{-1}_{\bm\Phi} \nabla_{\bm\Phi}\fscha = 0
\label{eq:preconditioner}
\eeq

It can be shown\cite{MauriPrecond,press2007numerical} that the condition number on the new problem defined by \eqname~\eqref{eq:preconditioner} is equal to
1 if $\bm A$ is the exact hessian matrix of $\fscha$. We can, therefore, use the analytic guess of the Hessian matrix $\bm A$ provided in \eqname~\eqref{eq:hessian}
to redefine the minimization algorithm. The SD algorithm on the new problem of \eqname~\eqref{eq:preconditioner} becomes:
\beq
\bm \Phi^{(n+1)} = \bm \Phi^{(n)} - \lambda  \bm A^{-1}_{\bm \Phi}\nabla_{\bm\Phi} {\mathcal F}({\vec{\mathcal R}}^{(n)}, \bm\Phi^{(n)}, \left\{\vec a_i\right\}),
\label{eq:sd}
\eeq
where $\lambda$ is the minimization step. Another advantage of using the auxiliary problem is that, if $\bm A$ is exact and $\fscha$ quadratic, the minimization arrives in the
minimum of the free energy in only one step with $\lambda = 1$. In a very similar way also the CG algorithm can be redefined for the auxiliary problem:
\begin{subequations}
\beq
\bm d^{(0)} = 0,
\eeq
\beq
\bm d^{(n+1)} = \bm A^{-1}_{\bm\Phi}\nabla_{\bm\Phi}\mathcal F^{(n+1)} + \frac{\nabla_{\bm\Phi}\mathcal F^{(n+1)} \bm A^{-1}_{\bm\Phi} \nabla_{\bm\Phi} \mathcal F^{(n+1)}}{\nabla_{\bm\Phi}\mathcal F^{(n)} \bm A^{-1}_{\bm\Phi} \nabla_{\bm\Phi} \mathcal F^{(n)}} \bm d^{(n)},
\eeq
\beq
\bm\Phi^{(n+1)} = \bm\Phi^{(n)}  - \lambda \bm d^{(n)}.
\label{eq:cg}
\eeq
\end{subequations}
Here, we omit the explicit dependence of the free energy $\nabla_{\bm\Phi}\mathcal F^{(n)} = \nabla_{\bm\Phi}\mathcal F({\vec {\mathcal R}}^{(n)}, \bm\Phi^{(n)}, \left\{\vec a_i\right\})$ for simplicity.

Since we can compute the Hessian matrix even of the fourth root problem, we can combine the two approaches of the nonlinear change of variable and the preconditioner 
to achieve a minimization constrained only on the positive defined $\bm\Phi$ with the smallest condition number.

\subsection{Hessian in the $\vec{\mathcal R}$ vector}\label{sec:prec:wyckoff}
The analysis on the minimization conducted so far investigates only the minimization problems faced with the $\bm \Phi$ parameter of the free energy. This is usually the
most problematic part of the minimization, since being a matrix, $\bm\Phi$ has much more degrees of freedom than the centroid positions. Furthermore, the centroid
positions are defined in the unit cell, while the force constant matrix is a supercell quantity. 
However, for generality,
it is very easy to provide an approximation also for the Hessian matrix of the free energy with respect to the $\vec{\mathcal R}$ variables.
This Hessian matrix $\bm A_{\vec{\mathcal R}}$ is the second derivative of the free energy\cite{Bianco2017}. However, since we are both neglecting mixed terms in the
Hessian, and we are taking an approximated Hessian also for the $\bm\Phi$ minimization, we chose:
\beq
\bm A_{\vec{\mathcal R}} = \bm\Phi_{SG}
\label{eq:wyckoff:h}
\eeq
This expression is correct if we keep $\bm\Phi$ fixed in the Hessian matrix derivative, and we compute $\bm A_{\vec{\mathcal R}}$ in $\bm\Phi(\vec{\mathcal R})$, 
where $\bm\Phi(\vec{\mathcal R})$ minimizes the free energy fixing the centroid positions to $\vec{\mathcal R}$. Moreover, \eqname~\eqref{eq:wyckoff:h} provides
a good preconditioner as it is always positive defined, and it does not require any additional computational effort to the algorithm.

The eigenvalues of $\bm\Phi$ are related to the square of the phonon frequencies for harmonic systems, therefore, we can approximate the condition number on the $\vec{\mathcal R}$ variables as
\beq
C_{\vec{\mathcal R}} \sim \left( \frac{\omega_{max}}{\omega_{\min}} \right)^2.
\label{eq:cn:r}
\eeq
This is not as pathological as the condition number seen on the $\bm \Phi$ minimization. However, we can introduce a preconditioner in the same way as described in
\secname~\ref{sec:precond} to handle easier minimization in low symmetry systems, as molecular crystals, where also many centroid degrees of freedom
 must be optimized, and 
the condition number~\eqref{eq:cn:r} can be of the order of $10^{6}$. 
Preconditioning also the $\vec{\mathcal R}$ variables allows one to have a dimensionless  step $\lambda$ for the minimization algorithms (\eqname~\ref{eq:sd} and~\ref{eq:cg}), with a clear advantage of reducing the human time necessary to optimize the two $\lambda$ steps for the $\bm \Phi$ and $\vec{\mathcal R}$ minimizations. 
We remark that the terms in the Hessian matrix obtained by the mixed derivatives in $\vec{\mathcal R}$ and $\bm \Phi$ are neglected.

\begin{figure}
\centering

\tikzstyle{line} = [draw, -latex']
\begin{tikzpicture}[auto, block/.style={rectangle, draw=black, text centered}, question/.style={diamond, draw = black, inner sep=0pt, text badly centered}]
\node [block, text width=8em] (init) {$\vec {\mathcal R_0}$, $\bm\Phi_0$ \\Compute the \\ preconditioner:\\ $\bm A_{\bm\Phi_0}$ $\bm A_{\vec{\mathcal R}_0}$};
\node [block, below of =init , node distance=1.5cm] (gen_ens) {Generate the ensamble}; 
\node [block, text width=8em, below of =gen_ens] (update) {Update the $\rho_I$ distribution};
\node [question, text width=4em, below of=update, node distance=2cm] (stat) {Is $\eta' < 0.6$?};
\node [block, text width=5em, below of=stat, node distance=2.2cm] (grad) {Compute $\vec{\mathcal R}$ and $\bm\Phi$ gradients};
\node [question, text width=6em, node distance=3cm, below of=grad] (conv) {Is the gradient  modulus comparable with its stochastic error?};
\node [block, text width=8em, node distance = 3cm, left of=grad] (minstep) {Minimization step\\ ${\vec{\mathcal R}}^{(n)} \rightarrow {\vec{\mathcal R}}^{(n+1)}$\\
$\bm\Phi^{(n)} \rightarrow \bm\Phi^{(n+1)}$\\See \figurename~\ref{fig:flowchart:nonlinear} and \\ Eqs.~\eqref{eq:sd}, \eqref{eq:cg}};
\node [question, text width=5em, node distance=4.5cm, below of=conv] (error) {Is the stochastic error sufficiently small?};
\node [block, node distance=2.5cm, below of=error] (done) {\bf DONE};
\node [block, node distance=3cm, text width=4em, right of=grad] (increase) {Increase the ensemble size};

\path [line] (init) -- (gen_ens);
\path [line] (gen_ens) -- (update);
\path [line] (update) -- (stat);

\node [left of=stat, node distance=2cm] (p1) {};
\node  {};
\path [line] (stat) -- node [near start] {yes} ($(stat) + (-2,0)$) -- ($(init) + (-2,0)$) -- (init);
\path [line] (stat) -- node [near start] {no} (grad);
\path [line] (grad) -- (conv);

\path [line] (conv) -| node [near start] {no} (minstep);
\path [line] (minstep) -- ($(update) +(-3, 0)$) -- (update);
\path [line] (conv) -- node [near start] {yes} (error);
\path [line] (error) -| node [near start] {no} (increase);
\path [line] (increase) -- ($(init) + (3,0)$) -- (init);
\path [line] (error) -- node [near start] {yes} (done);
\end{tikzpicture}

\caption{\small Flowchart of the new SSCHA implementation. The minimization step can be expanded by using the root4 algorithm introduced in \figurename~\ref{fig:flowchart:nonlinear}. In this case,  the preconditioner $\bm A_{\bm \Phi_0}$ should be replaced with  $\bm A_{\sqrt[4]{\bm\Phi_0}}$ in the initial step.
The minimization step is performed using the CG algorithm as long as the error is much greater than the stochastic noise, then the last steps are performed using SD.
This prevents error propagation in the conjugation due to the correlated noise introduced by the importance sampling reweighting procedure.
\label{fig:total:flowchart}}
\end{figure}
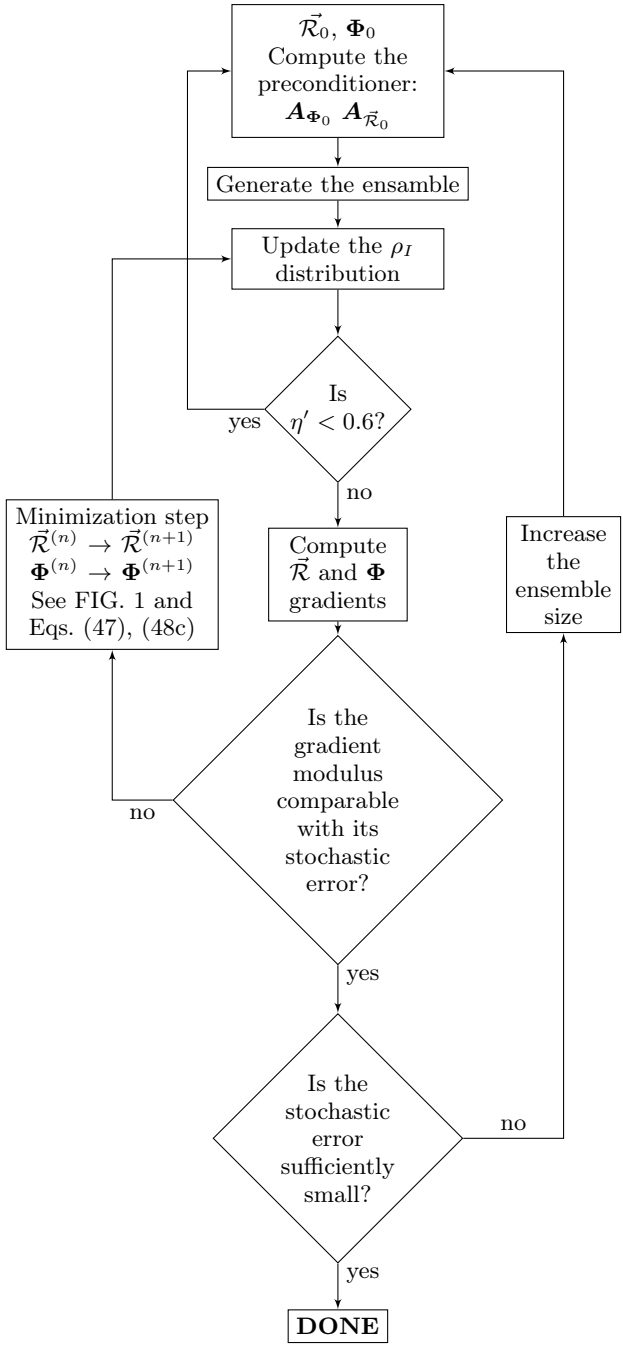

The new SCHA algorithm flowchart is shown in \figurename~\ref{fig:total:flowchart}.
 
% However, it is possible to compute this derivative even in the more general case of \eqname~\eqref{eq:wyckoff:h}. The gradient of the free energy respect to the atomic positions are the SCHA forces\cite{Errea2014}:
% \beq
% \frac{\partial\fscha}{\partial {\mathcal R}_a} = -\braket{f_a}_\rscha = - \int d^{3N}R \; f_a(\vec R) \rscha(\vec R)
% \eeq
% \beq
% \frac{\partial^2 \fscha}{\partial  {\mathcal R}_a\partial  {\mathcal R}_b} =  -\int d^{3N}R\; f_b(\vec R) \frac{\partial\rscha}{\partial {\mathcal R}_a} (\vec R)
% \eeq
% \beq
% \frac{\partial \rscha}{\partial {\mathcal R}_a} = \sum_{\beta = 1}^{3N} \Upsilon_{a\beta}u_\beta \rscha(\vec u),
% \eeq
% \beq
% \label{eq:wy:pre:prov}
% \frac{\partial^2  \fscha}{\partial {\mathcal R}_a\partial {\mathcal R}_b} = - \sum_{\beta = 1}^{3N}\Upsilon_{a\beta} \braket{u_\beta f_b}_\rscha.
% \eeq
% Summing and subtracting the SCHA harmonic forces inside the average in \eqname~\eqref{eq:wy:pre:prov} we get \eqname~\ref{eq:self:consistent} plus a correction:
% \beq
% \frac{\partial^2  \fscha}{\partial {\mathcal R}_a\partial {\mathcal R}_b} = \Phi_{ab} - \sum_{\beta = 1}^{3N} \Upsilon_{a\beta}\braket{u_\beta \delta f_b}_\rscha
% \eeq
% However, high energetic modes can force the minimization in the atomic positions to proceed much slower than the necessary in systems where light elements are abundant.

\section{Tests on ice XI ($\ch{H2O}$)}
\label{sec:forcefield}

In order to present the impressive enhancement in the minimization procedure obtained thanks to the combination of the preconditioning with the root representation, we report the calculation on the phase XI of ice. The difficulties of applying the SSCHA to this structure arise due to the presence of both hard covalent intra-molecular bonds and soft inter-molecular H-bonds resulting in a broad phonon spectrum.

Ice XI  is the proton ordered phase of common ice~\cite{Tajima1982} that is stable below 72~K. In this section we use a classical force-field that explicitly includes anharmonicity of the water molecule to compute energies and forces. The model is~\forcefield\cite{Pinilla_2014}.

\subsection{Stress tensor test}
\begin{figure}
\centering
\includegraphics[width=\columnwidth]{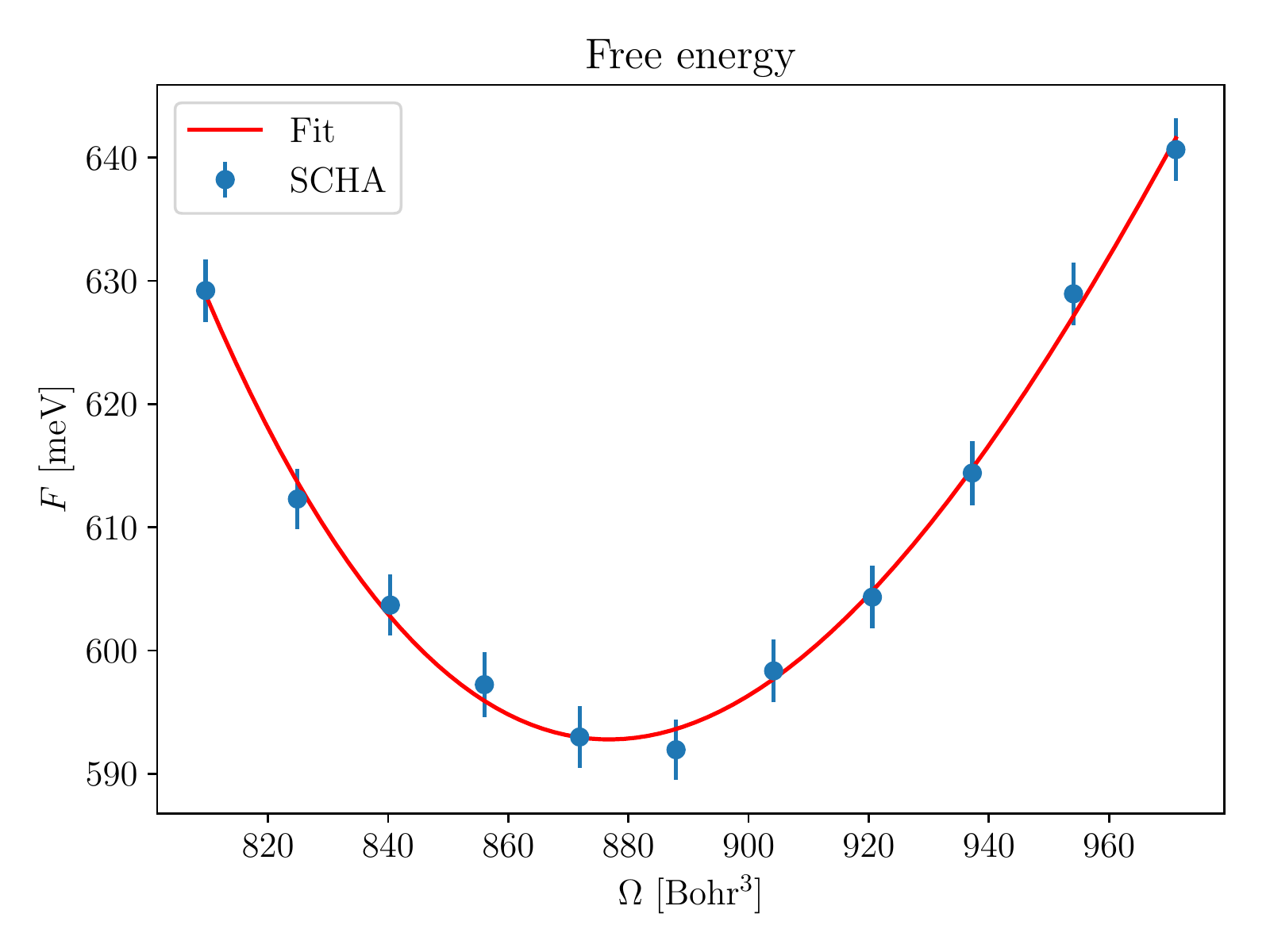}
\caption{\label{fig:FE} SCHA Free energy as a function of the volume. The unit cell is kept fixed, while only a isotropic scaling factor is considered.
The solid line represents a cubic fit. The simulation is performed at $T = \SI{100}{\kelvin}$.}
\end{figure}

\begin{figure}
\centering
\includegraphics[width=\columnwidth]{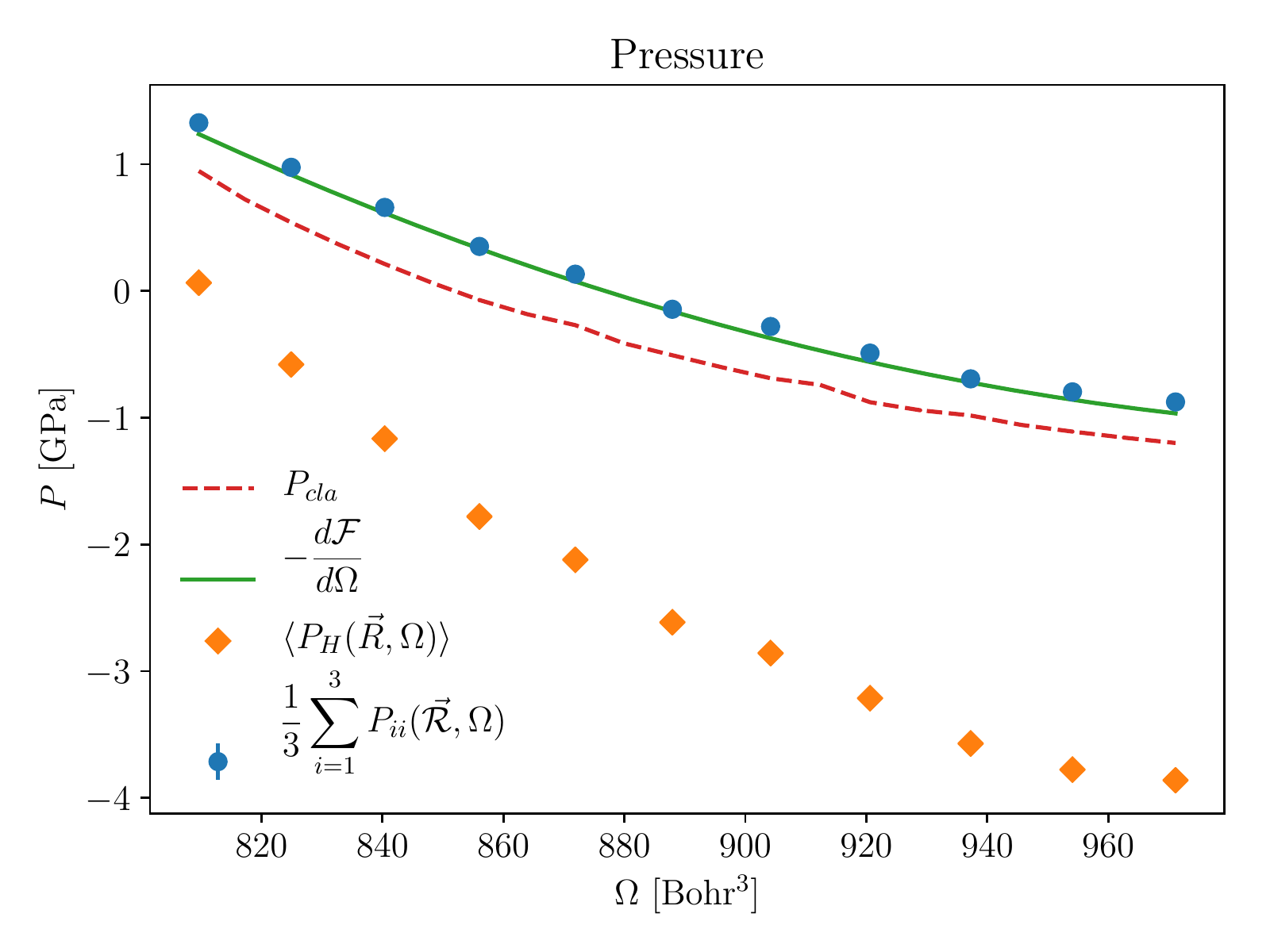}
\caption{\label{fig:P} (Color online) The figure compares the pressure computed with \eqname~\eqref{eq:p:final} (blue circles), the classical pressure $P_{cla}$ obtained neglecting thermal and quantum fluctuations (red dashed line), the average of the classical pressures over the SCHA ensemble (orange diamonds) and the analytical derivative of the Free energy fit reported in \figurename~\ref{fig:FE} (solid green line). The simulation is performed at $T = \SI{100}{\kelvin}$.}
\end{figure}

Here we test the anharmonic effects on the stress tensor with \forcefield.
\eqname~\eqref{eq:p:final} can be checked by performing the numerical derivative of the SCHA free energy at different volumes. 
In \figurename~\ref{fig:FE} we report the SCHA free energy as a function of the system volume, with a polynomial fit.
The cell is deformed with an isotropic expansion of the volume, so that the obtained pressure as the derivative of the
free energy versus the volume can be compared with 1/3 of the stress tensor trace of \eqname~\eqref{eq:p:final}. 
The fit on the SCHA free energy is then used to evaluate the pressure as a function of the volume:
\beq
P = - \frac{d\mathcal {F}}{d\Omega} = \frac 1 3 \sum_{\alpha=x,y,z}P_{\alpha\alpha}.
\label{eq:P:trace}
\eeq
In \figurename~\ref{fig:P} we compare the SCHA pressure obtained both as indicated in \eqname~\eqref{eq:P:trace} and as the opposite of the total derivative of the
free energy. The stochastic average of the stress tensors $\braket{P^H_{\alpha\beta}(\vec R, \left\{\vec a_i\right\}}_\rho$ is also reported, showing how the pressure cannot be considered as a physical
observable to be computed in analogy to what is done for general operators: $P_{\alpha\beta} \neq \braket{P^H_{\alpha\beta}}_\rho$. It is necessary to compute
it as the derivative of the free energy, as done in \eqname~\eqref{eq:p:final}. The pressure $P_{cla}$ without quantum effects at $T=0$ is also reported, and can be computed as 1/3 of the trace of the stress tensor in the classical equilibrium centroid positions:
\beq
P_{cla} = \frac 1 3 \sum_{\alpha = x, y, z} \frac{\partial V(\vec R_0, \left\{\vec a_i\right\} )}{\partial \varepsilon_{\alpha\alpha}},
\eeq
where $\vec R_0$ is defined as
\beq
\left.\frac{\partial V(\vec R, \left\{\vec a_i\right\})}{\partial \vec R}\right|_{\vec R = \vec R_0} = 0
\eeq
% The difference between this value and the anharmonic pressure 
% is the origin of the anomalous isotope volume effect in water\cite{Umemoto_2015,Pamuk_2012}. 

\subsection{Tests on the new minimization algorithm}
\begin{figure*}
\centering
\includegraphics[width=\textwidth]{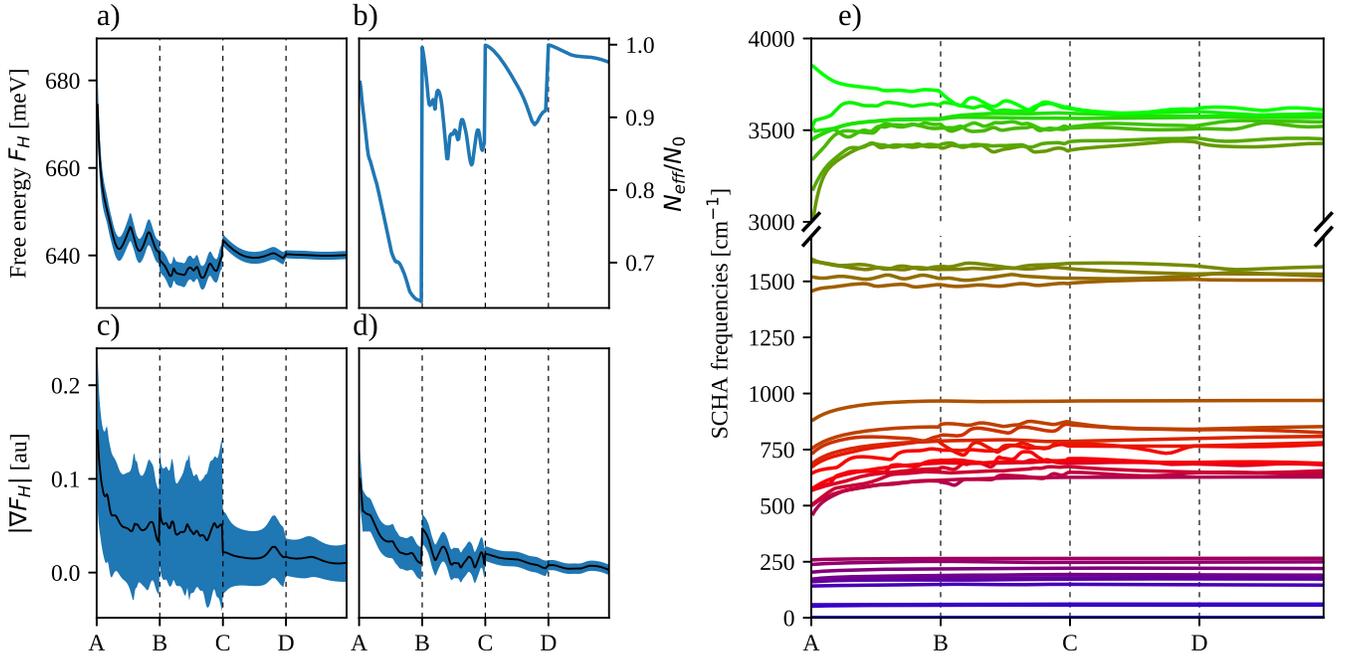}
\caption{\label{fig:minim}(Color online) Minimization progress starting from the harmonic result with the preconditioning linear change of variables. 
\textbf{a} The free energy. \textbf{b} The Kong-Liu effective sample size ratio, defined in \eqname~\eqref{eq:kl}.
\textbf{c} Modulus of the free energy gradient with respect to the dynamical matrix $\mathbf \Phi$. 
\textbf{d} Modulus  of the free energy gradient with respect to the centroids $\vec{\mathcal R}$.
\textbf{e} Frequencies obtained from the eigenvalues of the SCHA $\bm\Phi$ matrix, as they evolve during the minimization. Preconditioning uniformly converges all the
frequencies, achieving the final result  much faster. The \textbf{a-c-d} panels contains the stochastic error. For the two gradients the error is computed
as the norm of the error on each component of the gradient, to make it invariant with respect to the basis used to describe the $\bm\Phi$.}
\end{figure*}

A typical SCHA run with the precondition is reported in \figurename~\ref{fig:minim}. The $\rscha$ ensemble is re-extracted four times. The first two times ($A$ and $B$) 2500 configurations have been used, 10000 in the $C$ step, and 20000 in $D$. As clearly reported, the frequencies of the dynamical matrix converges uniformly to the final result, as we expect from the preconditioning, and we achieve a converged good result after only two steps.

\begin{figure}
\centering
\includegraphics[width=\columnwidth]{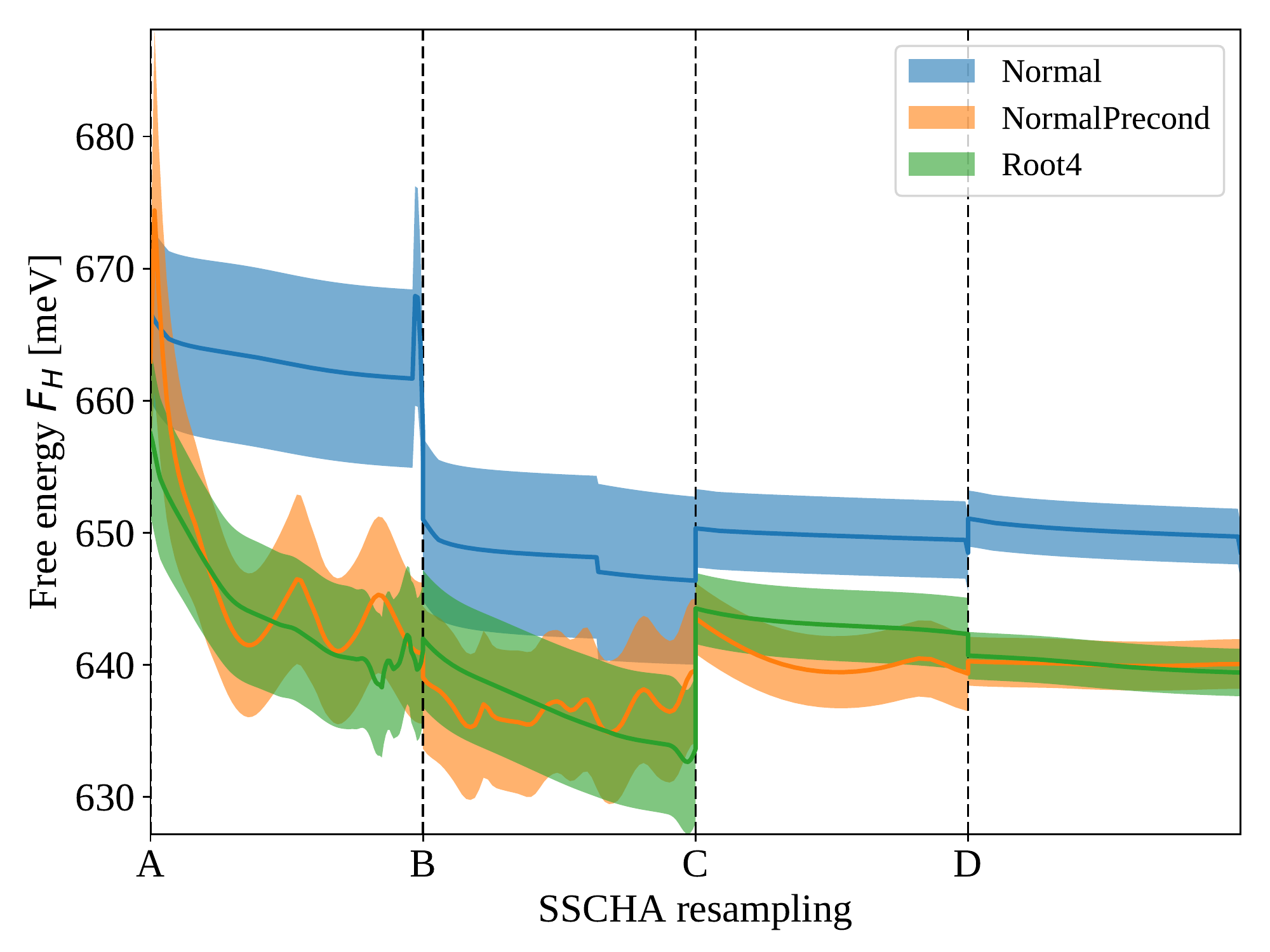}
\caption{\label{fig:comp}(Color online) Comparison between the different methods introduced here. The free energy is shown as a function of 
the number of configurations used for the stochastic evaluation together with its stochastic error. The first two calculations have 2500 configurations each. The third ($C$) is
with 10000. A final calculation is performed as a reference ($D$) with 20000 configurations to check the convergence of the previous ones.
As shown, the minimization without the preconditioning or the nonlinear change of variables is not able to get a converged result even using an overall of
35000 configurations, preventing the old SCHA to be used with any ``ab-initio'' technique in this kind of systems.
}
\end{figure}

The comparison of the performances between the nonlinear change $\mathbf \Phi \rightarrow \sqrt[4]{\mathbf \Phi}$ and the preconditioning is reported in \figurename~\ref{fig:comp}.
As a reference the SCHA run without the nonlinear change of variable and without preconditioning is also reported. The simulations are compared at $T = \SI{100}{\kelvin}$. 
It is clear that both two new methods greatly outperform the standard algorithm.
The harmonic dynamical matrix around the static equilibrium positions (neglecting quantum and thermal fluctuations) is used as a starting point, according to what is usually done in ``ab-initio'' calculations\cite{Errea2014,Errea2015,Nature2016}.
The \forcefield~harmonic dynamical matrix is close to the SCHA result, as seen by the low value of the free energy gradient with respect to $\bm\Phi$, compared
with its stochastic error, already in the first step of \figurename~\ref{fig:minim}. However, the standard minimization is not able to further minimize the system.

The success of the SCHA implementation on this force field paves the way to its systematic utilization for the study of water and any other complex
system with many degrees of freedom.%  However, as also shown in \figurename~\ref{fig:minim}, the anharmonic effect of the high energy frequencies, i.e. the stretching of the water molecules, affect the system hardening these modes, that is
% the opposite of what is known to happen experimentally.

In the next section we show the capabilities of the newly introduced method in a more realistic first-principles potential.

\section{Ab-initio simulation on ice XI} \label{sec:abinitio}
\begin{figure*}
\centering
\subfigure[Dispersion]{\includegraphics[width=0.68\textwidth]{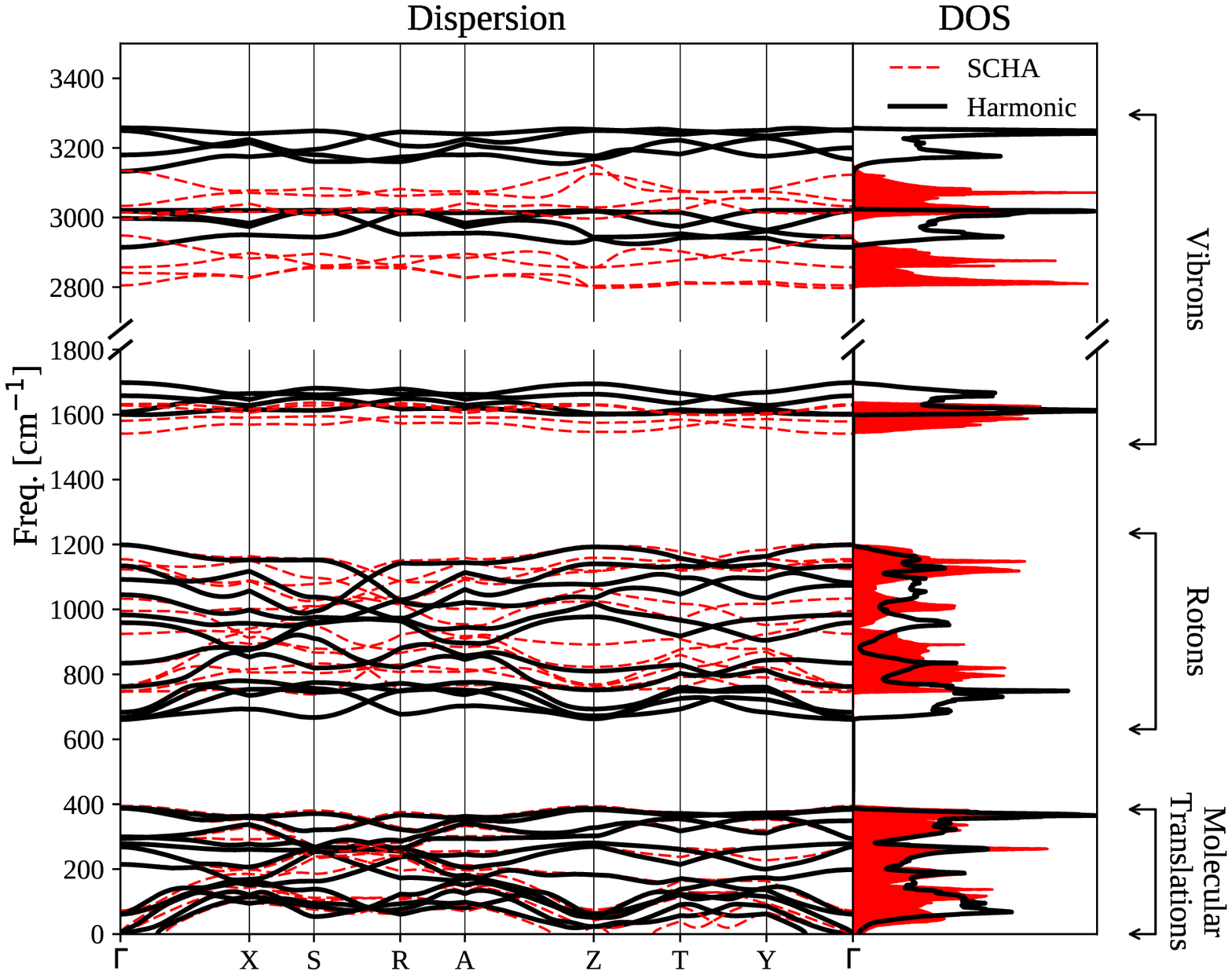}}
\subfigure[Unit cell]{\includegraphics[width=0.3\textwidth]{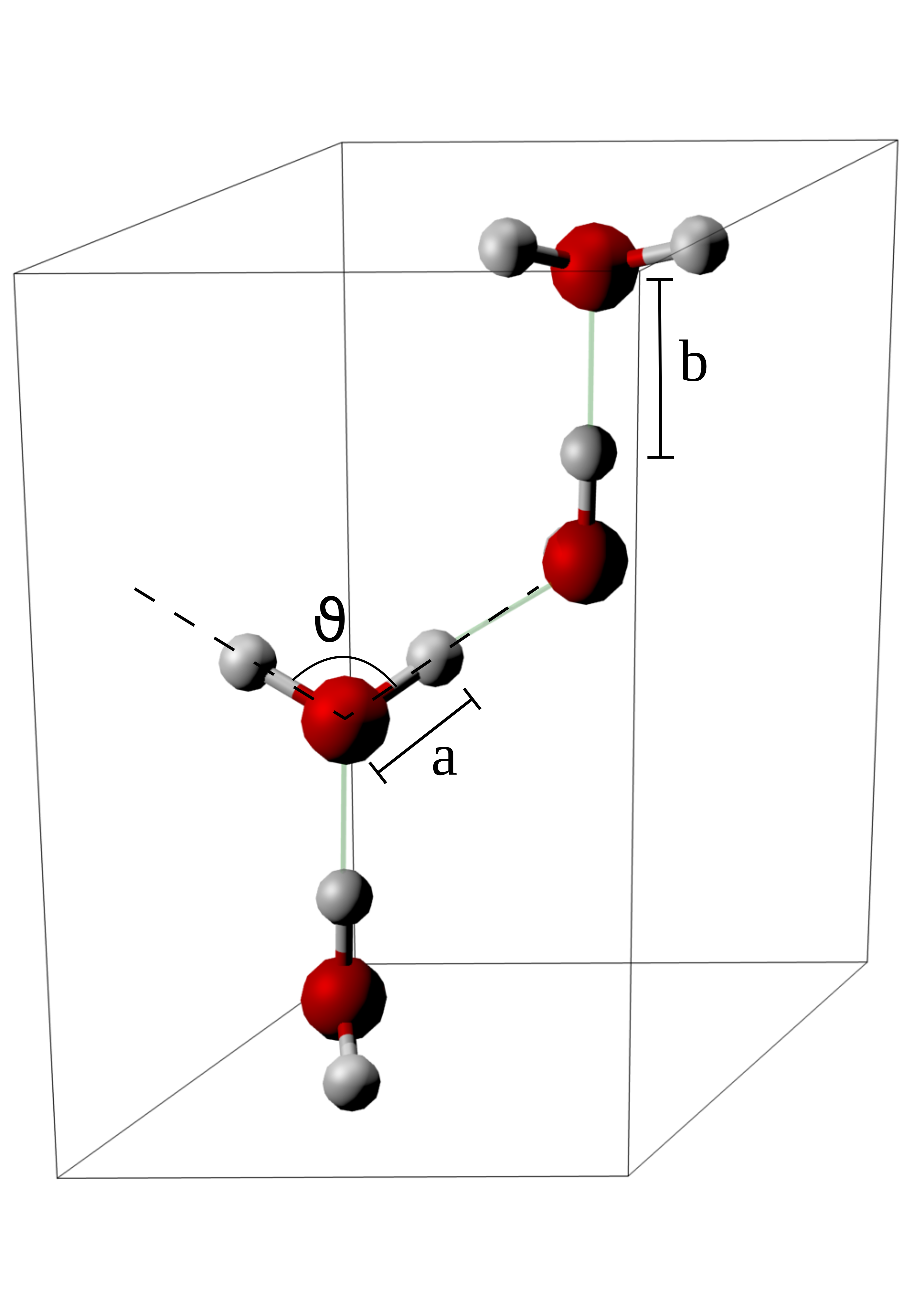}}
\caption{\label{fig:disp}(Color online) \textbf{a} Comparison between harmonic (black solid lines) and SCHA (red dashed lines) dispersion and density of states. 
\textbf{b} Unit cell of the ice XI structure. The parameters $a$, $b$ and
$\vartheta$ represent, respectively, the covalent OH bond, the hydrogen bond and the molecule angle. Their average value as a function of the temperature is reported in \tablename~\ref{table:centroids}.
}
\end{figure*}

Encouraged by the success of the new SCHA implementation on the \forcefield~force field, we report also the SCHA results on a realistic density-functional-theory (DFT) potential.
The converged SCHA phonon dispersion ($T = \SI{0}{\kelvin}$) is compared to the harmonic one 
in \figurename~\ref{fig:disp}. The calculation of energies and forces required to minimize the SCHA free energy, as well as the computation of the
harmonic dynamical matrix are performed \emph{ab-initio} with DFT, Perdew-Burke-Ernzerhof (PBE) exchange-correlation functional\cite{Perdew1996} and ultra-soft pseudo-potentials\cite{Vanderbilt1990} from the \emph{pslibrary}\cite{DalCorso2014}, as implemented in the  \textsc{QuantumESPRESSO} suite\cite{Giannozzi2009,Giannozzi2017}.
The SCHA dispersion is computed in the unit cell with 13000 overall configurations and a wave-function cutoff of $\SI{45}{\rydberg}$ ($\SI{360}{\rydberg}$  the charge density), then the difference between the harmonic and anharmonic dynamical matrices is extrapolated in a $3\times3\times3$ supercell, and the harmonic dispersion $\mathbf \Phi_0$ is added:
\beq
{\mathbf \Phi}^{(3\times 3 \times 3)} = \mathbf \Phi_{0}^{(3\times 3\times 3)} + \left({\mathbf \Phi}^{(1\times 1 \times 1)} - \mathbf \Phi_{0}^{(1\times 1\times 1)}\right)^{(3\times 3 \times 3)} .
\eeq
The harmonic phonon dispersion obtained interpolating the dynamical matrices converges already in a $2\times 2\times 2$ supercell, with a wave-function cutoff of $\SI{80}{\rydberg}$ ($\SI{640}{\rydberg}$ the charge density). The SCHA auxiliary dynamical matrix
$\bm\Phi$ is not directly related to the anharmonic phonon dispersion and, in general, a more sophisticated calculation is required to extract the real phonon frequencies in the SCHA approximation\cite{Bianco2017}. 
However, it is found\cite{Bianco2017} that the static phonon dispersion can be obtained as a perturbative series, whose leading order
is given by the $\bm\Phi$ matrix itself plus a ``bubble'' correction. It has been found in many system with hydrogen\cite{Nature2016,PaulattoPdH} that the ``bubble'' correction
is much lower than the $\bm\Phi$ contribution. As an explicative case, here we neglect this correction. Therefore, we report the anharmonic phonon dispersion and density of states approximated by directly interpolating the $\bm\Phi$ matrix after the SSCHA optimization in \figurename~\ref{fig:disp}. 

All phonon modes below $\SI{500}{\cm^{-1}}$ (molecular translations) result almost unaffected by the anharmonicity.
The two upper bands corresponding to symmetric and asymmetric stretching suffer a red-shift, together with the band around $\SI{1600}{\cm^{-1}}$ 
(molecular bending).
These modes are well described by molecular vibrons, and the observed red-shift is a general property of the water molecule\cite{Gillan_2016}.
Besides, the lowest part of the molecular rotations (the bands between $\SI{600}{\cm^{-1}}$ to $\SI{1200}{\cm^{-1}}$) are blue-shifted.
This  blue-shift of the lowest modes is indeed very interesting, since it involves inter-molecular modes. Such an effect of anharmonicity is typical of this
solid structure of ice, and cannot be predicted just studying anharmonicity in the isolated water molecule or the dimer.
 Moreover the Debye temperature of these bands is far above room temperature, invalidating the dispersion obtained with classical molecular dynamics, since zero point motion has a predominating role in these lattice oscillations.

Also the average atomic positions are affected as reported in \tablename~\ref{table:centroids}.  Here, the quantum fluctuations stretch slightly the water molecule, each covalent OH bond increases its length in almost a 0.7 \% and the molecule angle opens up in a 0.2\%. 
%Interestingly the covalent bond decrease its length with

\begin{table}
\begin{ruledtabular}
\begin{tabular}{lcccc}
 & Harmonic & $\SI{0}{\kelvin}$ & $\SI{150}{\kelvin}$ &$\SI{300}{\kelvin}$ \\
\hline
$a$ (OH-covalent) & $\SI{1.00835}{\angstrom}$ & $\SI{1.0159}{\angstrom}$ & $\SI{1.0148}{\angstrom}$ & $\SI{1.0131}{\angstrom}$  \\
% & 0\% & 0.51\% & 0.81 \% & 0.93\% \\
$b$ (H-bond) &$\SI{1.6769}{\angstrom}$ & $\SI{1.673}{\angstrom}$ & $\SI{1.674}{\angstrom}$ & $\SI{1.6765}{\angstrom}$ \\
$\vartheta$ (HOH angle) & $\SI{106.715}{\degree}$ & $\SI{106.950}{\degree}$ & $\SI{106.951}{\degree}$ & $\SI{106.886}{\degree}$
%H-bond asym. & 0.048\% & 0.69 \% & 0.52\% & 0.17\%
\end{tabular}
\end{ruledtabular}
\caption{\label{table:centroids}Anharmonic effects on the crystal structure predicted by the DFT-PBE at three temperatures. The average intramolecular OH distance (covalent bond), the average H-bond distance and the water molecule mean angle, as reported in \figurename~\ref{fig:disp}.}
\end{table}

% These results are represented by the radial distribution function  of the OH and HH distances, reported in \figurename~\ref{fig:gr}. The 
% radial distribution function $g(r)$ is defined as:
% \beq
% g_{XY}(r) = \frac{1}{N_X N_Y} \sum_{i = 1}^{N_X} \sum_{j = 1}^{N_Y} \braket{\delta(|\vec r_{ij}| - r)}_\rho,
% \eeq
% where $N_X$ is the number of $X$ particles in the system.
% The $g(r)$ can be computed straightforwardly in the SCHA framework even at $T = 0$ (full quantum), thanks to the analytical expression of the SCHA density matrix.

% \begin{figure*}
% \centering
% \subfigure[OH]{
% \includegraphics[width=0.45\textwidth]{FigureGR/gr.eps}}
% \subfigure[HH]{
% \includegraphics[width=0.45\textwidth]{FigureGR/gr_HH.eps}}
% \caption{\label{fig:gr}(Color online) Radial distribution functions of the quantum vibrational ensemble at $T = 0$ in the harmonic and in the SCHA.
% Anharmonic effects stretch the covalent bond of the water molecule and shorten the hydrogen bond.}
% % The python script that computes the figure for the g(r) is located into Documenti/ICE/StudyVolume/PBE/V_814.*/minimizations
% \end{figure*} 

Even if the anharmonic molecular stretch can seem negligible compared to what it has been predicted for
a high-pressure molecular phase of hydrogen\cite{Borinaga2016}, a difference  of 1\% in the OH covalent bond has a great contribution to the energy. As a test, the SCHA average structure can be used for a classical DFT calculation, where the classical pressure is found to be $\SI{1}{\giga\pascal}$ lower (negative) than its value in the equilibrium positions, suggesting that the anharmonic relaxation of the centroid positions may significantly affect the pressure, and, consequently, the equilibrium volume. 

\begin{figure*}
\centering
\subfigure[]{
\includegraphics[width=0.49\textwidth]{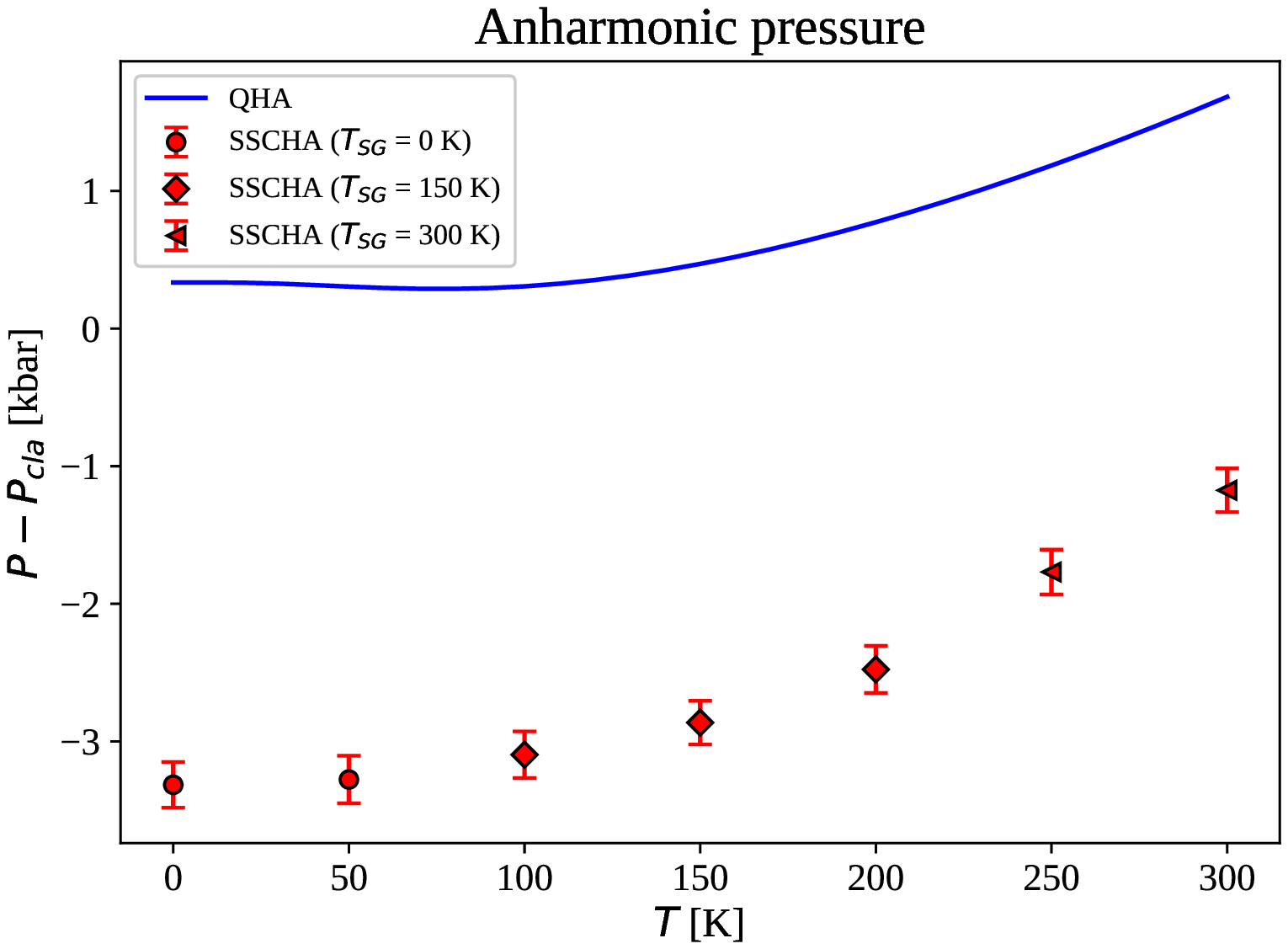}}
\subfigure[]{
\includegraphics[width=0.49\textwidth]{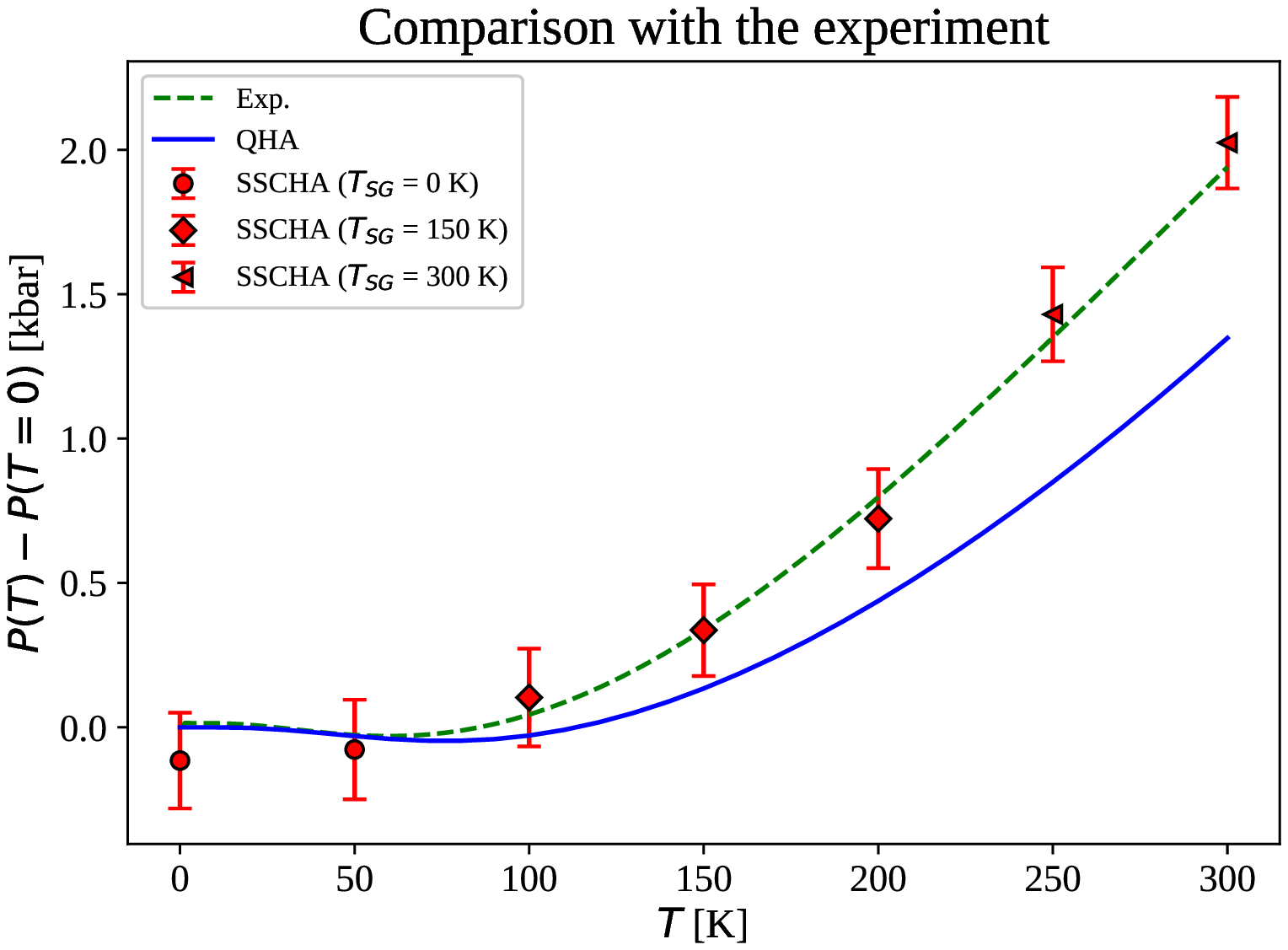}}
\caption{\small (Color online) \textbf{a:} Comparison between the QHA and the SSCHA pressure as a function of the temperature at fixed volume. The zero value of the pressure is $P_{cla}$, obtained as 1/3 of the stress tensor trace at $T= 0$ neglecting quantum fluctuations. \textbf{b:}
Comparison between QHA and SSCHA simulations and the fit of the experimental results from Ref.[\onlinecite{FortesExp}] (ice Ih). 
The SSCHA results are computed at temperatures of $\SI{0}{\kelvin}$, $\SI{150}{\kelvin}$ and $\SI{300}{\kelvin}$ with 40000 stochastic configurations. The other temperatures reported have been obtained through reweighting\cite{Errea2014}, therefore their stochastic error is correlated.
\label{fig:qhavssscha}} 
\end{figure*}

The stress tensor calculation can be used to optimize the unit cell considering both thermal and quantum  effects. The most advanced calculations to include these effects without involving PIMD in water
have been performed within the  quasi-harmonic approximation (QHA)\cite{Umemoto_2015,Pamuk_2012}.
In this scheme the total pressure is obtained expanding the BO energy surface as a quadratic function around its minimum at each volume. Then the exact free energy of the approximated BO surface can be computed analytically:
\begin{align}
\mathcal F_{QHA}(\vec R_c, \left\{\vec a_i\right\}) &= V(\vec R_c, \left\{\vec a_i\right\})   \nonumber
+ \sum_{\mu = 1}^{3N} \bigg[\frac{\hbar \tilde\omega_\mu(\vec R_c, \left\{\vec a_i\right\})}{2} +\\ 
&+ \frac 1\beta\ln\left(1 - e^{-\beta\hbar\tilde\omega_\mu(\vec R, \left\{\vec a_i\right\})}\right)\bigg] ,
\end{align}
where $\tilde\omega_\mu$ are the harmonic frequencies of the BO surface. The QHA free energy $F_{QHA}$ is obtained minimizing the functional $\mathcal F_{QHA}$ at fixed 
volume and temperature:
\beq
F_{QHA}(T, \left\{\vec a_i\right\}) = \min_{\vec R_c} \mathcal F_{QHA}(T, \vec R_c, \left\{\vec a_i\right\}).
\label{eq:fe:qha}
\eeq
The QHA pressure is obtained by differentiating the free energy with respect to a uniform volume deformation:
\beq
P_{QHA} = -\frac{dF_{QHA}}{d\Omega}
,
\label{eq:p:qha}
\eeq
In complex systems with many degrees of freedom, like in ice, the minimization in \eqname~\eqref{eq:fe:qha} is computationally very expensive, since it requires the calculation
of the gradient of the free energy (that depends on the harmonic dynamical matrix) with respect to any possible atomic displacement. This involves the calculation of a third order
derivative of the BO total energy for each minimization step\cite{Lazzeri_2002}. 
Differences and analogies of QHA and SCHA approaches are discussed in appendix~\ref{app:QHA:vs:SCHA}.
No QHA implementation with the full atomic coordinates relaxation has been 
performed in $\ch{H2O}$ systems up to now, and the QHA free energy is approximated with $\vec R_c = \vec R_0$:  the minimum of the BO energy.
The pressure in \eqname~\eqref{eq:fe:qha} is computed numerically taking finite differences between the QHA free energies at several volumes.
A more convenient way to compute the QHA pressure is to consider the harmonic frequencies as a linear function of the volume:
\beq
\tilde \omega_k(\Omega) = \tilde \omega_k(\Omega_0) \left[ 1 - \frac{\Omega - \Omega_0}{\Omega_0}\gamma_k\right],
\eeq
where the $\gamma_k$ are the Gr\"uneisen parameters.
Then the QHA pressure can be easily obtained at any temperature:
\beq
P_{QHA} = P_H(\Omega) - \sum_{\mu = 1}^{3N} \frac{\hbar\omega_\mu \gamma_\mu}{2\Omega}  \frac{1}{\tanh\left(\frac{\beta\hbar\omega_\mu}{2}\right)}.
\eeq
The comparison between QHA and the SSCHA pressure calculations as a function of temperature and is reported in panel $a$ of \figurename~\ref{fig:qhavssscha}. 

The experimental fit on the elastic bulk modulus and the volume expansivity have been used to compare the QHA and SSCHA pressures at fixed volume. 
As clearly shown, the QHA pressure is shifted by about $\SI{4}{\kilo\bar}$ with respect to the SSCHA result. This is two times bigger than the whole pressure
range between $\SI{0}{\kelvin}$ and $\SI{300}{\kelvin}$. In panel $b$ of \figurename~\ref{fig:qhavssscha} the comparison between the QHA, the SSCHA and the experimental
results is reported. All the pressure are shifted respect their zero temperature value. The SSCHA zero temperature pressure has been obtained by fitting the SSCHA points with the experimental curve. The experimental data have been obtained from the fit reported in Ref.[\onlinecite{FortesExp}].

The QHA grasps the qualitative behaviour of the pressure, including the low temperature negative thermal expansion\cite{gupta2018phonons}, but it deviates from the experiments at temperatures above $\SI{80}{\kelvin}$. This effect has been associated with the entropy contribution of the proton disorder of the ice Ih respect to the ice XI,
that is not accounted in the simulations\cite{gupta2018phonons}. However, the SSCHA result is in perfect agreement with the experiments, suggesting that the underestimation of the pressure at high temperature can be simply explained as a failure of the QHA. This indicates that anharmonic effects beyond QHA play an important role in reproducing
the physical properties of ice at temperatures above $\SI{80}{\kelvin}$.

Another interesting feature of ice at low temperature is the anomalous volume isotope effect: the $\ch{D2O}$ equilibrium volume is bigger than the $\ch{H2O}$ one.
This effect has been recently studied within the QHA\cite{Pamuk_2012,Umemoto_2015}.
In particular, Pamuk \emph{et. al.}\cite{Pamuk_2012} showed how the QHA result systematically overestimates this effect with several DFT functional
but it depends slightly on the chosen functional, on the difference between ice Ih and XI, and on the $q$-grid interpolation.
The experimental\cite{Rottger1994} difference between the two volumes at $\SI{10}{\kelvin}$ is about a 0.09\%, 
while the difference between the QHA equilibrium volume and the SSCHA one is a 1.8\%. Therefore, the isotope volume effect is a tiny correction respect to
the ZPE contribution on the equilibrium volume and the difference between the SSCHA and the QHA.

\section{Conclusions}\label{sec:conclusions}
The study of quantum anharmonic effects in complex crystals with lots of degrees of freedom, e.g., molecular crystals, is a major challenge that impacts in many domains of physics and chemistry, including high pressure phases of hydrogen, water anomalies, thermoelectric materials, charge-density waves, ferroelectrics, multiferroics, and so on. In this work, we derive a new expression for the anharmonic
contribution to the stress tensor in the SCHA theory. This correction is very important for accurate pressure estimations and phase-diagram computations and paves the way for the isobaric unit cell relaxation. We further improved the stochastic implementation of the SCHA theory to
apply it in complex crystals with a large number of degrees of freedom. This aim has been achieved thanks to a clever preconditioning on the free energy minimization algorithm,
based on an analytical guess of the Hessian matrix of both the force-constant matrix and the central nucleus positions, and with a nonlinear change of variables that
restricts the space of allowed dynamical matrices only to the positive defined ones. 

The new algorithm is benchmarked with the phase XI of ice, the proton-ordered phase of common ice, a prototype molecular crystal.
The quantum ZPM and anharmonicity are proven to affect the phonon dispersion both in the molecular and in the intermolecular modes. Also, the $OH$ and $HH$ bound distances are slightly affected by anharmonicity.
The importance of the non-perturbative SCHA contribution to the pressure in this system has been benchmarked
%, and compared with what obtained with other state-of-art harmonic based technique (QHA). The both qualitative and quantitative difference in the results arises both from the dynamical matrix relaxation, and from the huge effect of anharmonicities on the nuclei positions, not considered in harmonic approaches. This result make us wonder whether the interpretation of water quantum anomalies based only on QHA calculation, without relaxing the crystal structure, are valid. 
in~\forcefield~and calculated with \emph{ab-initio} DFT PBE, where the quantum fluctuations at $\SI{0}{\kelvin}$ are shown to affect the equilibrium volume by
a 1.8\%.
 
The thermal expansion of the system has been computed both within the QHA and the SSCHA. The QHA is found to miscalculate both the ZPM contribution to the equilibrium volume (with a wrong sign) and the effect of the thermal fluctuations at temperatures above $\SI{80}{\kelvin}$.
The latter discrepancy was associated with the proton disorder of phase Ih of ice. However, we found the SSCHA to be correct this effect, and to exhibit an excellent agreement with the experiments, unveiling that the QHA fails in accounting the anharmonicity of ice.

The new  cell-relaxation and stress calculation here introduced paves the way to a refreshed quantitative and accurate study of anharmonic effects on water, like the anomalous 
isotope volume, the equilibrium isotope fraction, the negative thermal expansion, and the high-pressure phase diagram.
More generally, the introduced stress tensor derivation and the improved minimization algorithm make the SSCHA an efficient method to calculate quantum and thermal  anharmonic effects on complex systems with many degrees of freedom.

\begin{acknowledgments}
IE acknowledges financial support from the Spanish Ministry of Economy and Competitiveness (Grant No. FIS2016-76617-P).
We acknowledge the CINECA award under the
ISCRA initiative (Grant HP10BLTB9A), PRACE, IDRIS, CINES
and TGCC under the EDARI project A0030901202 for
the availability of high performance computing resources
and support. 
\end{acknowledgments}

\appendix

\begin{widetext}

\section{Stress tensor derivation}
\label{app:stress1}
To easily compute the derivative of the SCHA free energy functional with respect to the strain tensor is convenient to use the formalism introduced by Bianco \emph{et al}\cite{Bianco2017}. The average of a generic observable can be written as:
\beq
\braket{O}_\rscha = \sqrt{\det\left(\frac{\bm\Upsilon}{2\pi}\right)} \int O(\vec{ \mathcal R} + \vec u, \{\vec a_i\})\exp\left( - \frac 1 2 \vec u\,\bm \Upsilon \vec u\right) \, d^{3N} u,
\eeq
In order to normalize the Gaussian integral a change of variable can be applied, so that:
\beq
u_s^\alpha = \sum_\mu {J_\mu}_s^\alpha y_\mu \qquad {J_\mu}_s^\alpha = \frac{{e_\mu}_s^\alpha}{\sqrt {M_s}}\sqrt{\frac{\hbar(1 + n_\mu)}{2\omega_\mu}}
\qquad
\left(\bm \Upsilon^{-1}\right)^{\alpha\beta}_{st} = \sum_\mu {J_\mu}_s^\alpha {J_\mu}_t^\beta.
\eeq
Then we have:
\beq
\braket{O}_\rscha = \int O(\vec {\mathcal R} + \mathbf J \vec y, \{\vec a_i\}) \left[dy\right] \qquad \left[dy\right]= \prod_{\mu = 1}^{3N} \frac{\exp\left(\frac{-y_\mu^2}{2}\right)}{\sqrt{2\pi}} dy_\mu.
\eeq
Since we are deriving the $\mathcal F_{\vec{\mathcal R}}$ functional (\eqname~\ref{eq:func:f}), the Hellman-Feynman (HF)  theorem allows us to neglect the changes introduced by the strain on the dynamical matrix. Only $\vec{\mathcal R}$ is affected by the deformation, according to \eqname~\eqref{eq:strain:def}.
Therefore we have:
\beq
\frac{d\braket{O(\vec R, \left\{\vec a_i\right\})}_{\rho_{\mathcal R}}}{d\varepsilon_{\alpha\beta}} = \left.\frac{\partial\braket{O(\vec R,\left\{\vec a_i\right\} )}_\rscham}{\partial \varepsilon_{\alpha\beta}}\right|_{\bm\Phi = \bm\Phi(\vec{\mathcal R})},
\eeq
\beq
\frac{\partial \braket{O}_\rscham}{\partial\varepsilon_{\alpha\beta}}  = \frac{\partial}{\partial\varepsilon_{\alpha\beta}} \int O\left(\vec{\mathcal R} (\bm\varepsilon) + \bm J\vec y, \left\{\vec a_i(\varepsilon)\right\}\right)\left[ dy\right] = \int \left(\sum_{s\gamma} \frac{\partial O}{\partial {  R}_s^\gamma} \frac{\partial {\mathcal R}_s^\gamma}{\partial\varepsilon_{\alpha\beta}} + \sum_{i\gamma} \frac{\partial O}{\partial a_i^\gamma} \frac{\partial a_i^\gamma}{\partial\varepsilon_{\alpha\beta}}\right) \left[dy\right],
\eeq
Note that the observable $O(\vec R)$ is derived respect to its argument, i.e. the atom positions in the ensemble configuration $\vec R$, not the centroid position $\vec {\mathcal R}$.
This happens because the $\vec{\mathcal R}(\bm\varepsilon)$ appears linearly in the configuration position of $O$ after the change of variable.
\beq
\frac{\partial \braket{O}_\rscham}{\partial\varepsilon_{\alpha\beta}}  =  \frac 12\sum_s \left({\mathcal R}_s^\beta\braket{\frac{\partial O}{\partial R_s^\alpha}}_\rscham +{\mathcal R}_s^\alpha \braket{\frac{\partial O}{\partial R_s^\beta}}_\rscham \right) + \braket{\sum_{i\gamma} \frac{\partial O}{\partial a_i^\gamma} \frac{\partial a_i^\gamma}{\partial \varepsilon_{\alpha\beta}}}_\rscham.
\label{eq:average}
\eeq
The free energy functional is:
\beq
{\mathcal F}_{\vec {\mathcal R}} = F_{\mathbf \Phi(\vec{\mathcal R})} + \braket{V - \mathcal V_{\vec{\mathcal R}, \mathbf\Phi(\vec{\mathcal R})}}_{\rho_{\vec{\mathcal R}}},
\eeq
where $\mathbf \Phi(\vec{\mathcal R})$ is the dynamical matrix that minimizes $\fscha$ fixing the average atomic positions.
The first term, $F_{\mathbf \Phi(\vec {\mathcal R})}$ is an explicit function only of the SCHA dynamical matrix, therefore does not contribute to the derivative.
The latter average can be derived thanks to \eqname~\eqref{eq:average}:
\beq
\frac{\partial \braket{V}_\rscham}{\partial \varepsilon_{\alpha\beta}} =- \frac 1 2 \sum_{s = 1}^N \left({\mathcal R}_s^\beta \braket{f_s^\alpha}_\rscham + {\mathcal R}_s^\alpha \braket{f_s^\beta}_\rscham\right) -\Omega \braket{P_{\alpha\beta}^H}_\rscham,
\label{eq:pbo}
\eeq
where $P_{\alpha\beta}^H$ is the BO stress tensor. In fact the last term of \eqname~\eqref{eq:average} is the average of the derivatives of the BO energy when the
strain is applied to the unit cell. The ``harmonic'' term can be computed in a similar way:
\beq
\left.\frac{\partial \braket{\mathcal V_{\vec{\mathcal R}, \bm\Phi}}_\rscham}{\partial\varepsilon_{\alpha\beta}}\right|_{\bm\Phi} = 
\frac 12\sum_s \left({\mathcal R}_s^\beta\braket{\frac{\partial \mathcal V_{\vec{\mathcal R}, \bm\Phi}}{\partial R_s^\alpha}}_\rscham +{\mathcal R}_s^\alpha \braket{\frac{\partial \mathcal V_{\vec{\mathcal R}, \bm\Phi}}{\partial R_s^\beta}}_\rscham \right) +
 \braket{\frac{\partial \mathcal V_{\vec{\mathcal R}, \bm\Phi}}{\partial \varepsilon_{\alpha\beta}}} .
\eeq 
In the same way as done for the BO energy surface, it is possible to introduce the harmonic  stress tensor as
\beq
P^{\mathcal H}_{\alpha\beta} = - \frac 1 \Omega \braket{\frac{\partial \mathcal V_{\vec{\mathcal R}, \bm\Phi}}{\partial\varepsilon_{\alpha\beta}}}_\rscham= \frac {1}{2\Omega}\sum_s \braket{{f_{\mathcal H}}_s^\alpha u_s^\beta + {f_{\mathcal H}}_s^\beta u_s^\alpha}_\rscham = 
- \frac{1}{\Omega}  \sum_{\mu = 1}^{3N}\sum_{s = 1}^N \frac{\hbar \omega_\mu}{2\tanh\left(\frac{\beta\hbar\omega_\mu}{2}\right)} {e_\mu}_s^\alpha {e_\mu}_s^\beta,
\eeq
\beq
\frac{\partial \braket{\mathcal V}_\rscham}{\partial \varepsilon_{\alpha\beta}} =  - \frac 1 2 \sum_{s = 1}^N \left({\mathcal R}_s^\beta \braket{{f_{\mathcal H}}_s^\alpha}_\rscham + {\mathcal R}_s^\alpha \braket{{f_{\mathcal H}}_s^\beta}_\rscham\right)  
- \Omega P_{\alpha\beta}^{\mathcal H}
% - \frac 1 2\sum_s \braket{{f_{\mathcal H}}_s^\alpha u_s^\beta + {f_{\mathcal H}}_s^\beta u_s^\alpha}_\rscham
.
\eeq
The first term is zero (the harmonic forces $\vec f_{\mathcal H}$ are odd, while the probability distribution $\rscham$ is even). However, we keep it as it helps increasing the numerical accuracy\cite{Errea2014}, as we can combine it with \eqname~\eqref{eq:pbo} to exploit the correlation between $f_s^\alpha$ and ${f_{\mathcal H}}_s^\alpha$
to reduce the statistical noise on the average. In a pure harmonic crystal also the quantities $P_{\alpha\beta}^H$ and $P_{\alpha\beta}^{\mathcal H}$ are correlated. Therefore, the final expression of the pressure can be written as follows
\beq
P_{\alpha\beta} = \braket{P_{\alpha\beta}^H - \frac{1}{2\Omega} \sum_{s = 1}^N \left({f_{\mathcal H}}_s^\alpha u_s^\beta + {f_{\mathcal H}}_s^\beta u_s^\alpha\right)}_\rscham + \frac{1}{2\Omega} \sum_{s = 1}^N \left({\mathcal R}_s^\beta \braket{f_s^\alpha - {f_{\mathcal H}}_s^\alpha}_\rscham + {\mathcal R}_s^\alpha \braket{f_s^\beta - {f_{\mathcal H}}_s^\beta}_\rscham\right).
\eeq
The last term is zero zero if the free energy has been minimized also with respect to the $\vec{\mathcal  R}$ variables (as the average of the BO forces is the SCHA force acting on each atom, it is zero in the equilibrium).

\section{Detailed calculation for the hessian matrix}
\label{app:hessian}

The real and trial classical forces acting on each configuration identified by the displacements $\vec u$ are:
\begin{subequations}
\beq
f_s^{\alpha} = -\frac{\partial V}{\partial u_s^{\alpha}} = -\sum_{t\beta}K^{\alpha\beta}_{st} u_t^{\beta},
\eeq
\beq
{f_\mathcal H}_s^{\alpha} = -\frac{\partial\mathcal V_{\vec{\mathcal R}, \mathbf\Phi}}{\partial u_s^{\alpha}} = - \sum_{t\beta} \Phi^{\alpha\beta}_{st} u_t^{\beta}.
\eeq
\end{subequations}
Defining $\vec{\delta f} = \vec f -\vec  f_{\mathcal H}$ we have
\beq
\braket{\delta f_s^{\alpha} u_t^{\beta}}_{\rho_{\mathcal H}}  = - \sum_{n\eta}(K_{sn}^{\alpha\eta} - \Phi_{sn}^{\alpha\eta})\braket{u_n^{\eta}u_t^{\beta}}_{\rho_{\vec {\mathcal R}, \mathbf\Phi}}. 
\label{eq:prova}
\eeq
From now on, we drop the subscript $\rho_{\mathcal {\vec R}, \mathbf{\Phi}}$ for each average, and consider all the averages computed with respect to the trial density matrix. We further
simplify the notation, introducing one index for each Cartesian and atomic coordinate, so $v_s^\alpha \rightarrow v_a$. In this new notation \eqname~\eqref{eq:prova} reads:
\beq
\braket{\delta f_a u_b}  =  - \sum_{c = 1}^{3N}(K_{ac} - \Phi_{ac})\braket{u_cu_b}. 
\eeq
The average of the product between two displacements of a Gaussian distributed variable is the covariance between the two displacements (\eqname~\ref{eq:rho}):
\beq
\braket{u_c u_b} = \left(\mathbf \Upsilon^{-1}\right)_{cb} =\frac{1}{\sqrt {M_c M_b}} \sum_{\nu = 1}^{3N} e_\nu^{c} e_\nu^{b} a_\nu^2,
\eeq
where we introduce the mode length $a_\mu$:
\beq
a_\mu = \sqrt{\frac{\hbar}{2\omega_\mu} \left(1 + 2 n_\mu\right)} .
\eeq
The gradient of the SCHA free energy functional with respect to $\mathbf \Phi$ is\cite{Errea2014}:
\beq
  \nabla_{\mathbf \Phi}\mathcal F_{\vec{\mathcal R}, \mathbf\Phi}  = - \sum_{ab \mu}
  \sqrt\frac{M_a}{M_b} \left(e_\mu^{b}\nabla_{\mathbf \Phi} \ln a_\mu +  \nabla_{\mathbf \Phi} e_\mu^{b}\right)
  e_\mu^{a}\braket{ \delta f_bu_a} .
\label{eq:gradient}
\eeq
Substituting the explicit expression of the forces we have:
\begin{align}
  \nabla_{\mathbf\Phi} \fscha  =  \sum_{a b c \mu \nu} &\nonumber
\left(K_{ac} - \Phi_{ac}\right)\left(e_\mu^{a} \nabla_{\mathbf \Phi} \ln a_\mu + \nabla_{\mathbf \Phi} e_\mu^{a}\right)\times\\
&\times \frac{e_\mu^{b}e_\nu^{c}e_\nu^{b}a_\nu^2 }{\sqrt{M_cM_a}}.
\label{gradient:reduced}
\end{align}
It is clear from \eqname~\eqref{gradient:reduced} that in the minimum $ \mathbf \Phi = \mathbf K$.
Therefore, it is convenient to compact all the other terms into a symbol:
\beq
\label{grad:fh:harmonic}
\frac{\partial \fscha}{\partial \Phi_{cd}} = \sum_{ab}\left(K_{ab} - \Phi_{ab}\right){ \mathcal L}_{abcd}.
\eeq
Here $\bm{\mathcal L}$ is a 4-rank tensor. Since we sum on all $a$ and $b$ indexes and $\bm{\mathcal L}$ 4-rank tensor multiplies a symmetrical matrix, 
it is convenient to recast it into a symmetrical form:
\beq
\label{l:definition}
{L}_{abcd} = \sum_{k,\mu\nu} \left(e_\mu^{a} \frac{\partial \ln a_\mu}{\partial\Phi_{cd}} +  \frac{\partial e_\mu^{a}}{\partial\Phi_{cd}}\right)e_\mu^{k}e_\nu^{b}e_\nu^{k}a_\nu^2,
\eeq
\beq
{\mathcal L}_{abcd} = \frac{\mathcal P_{ab}}{\sqrt{M_aM_b}}\frac{  L_{abcd} + L_{bacd}}{2},
\eeq
\beq
\mathcal P_{ab} = \sqrt 2(1 - \delta_{ab}) + \delta_{ab},
\eeq
\beq
\label{eq:mat:l}
{\mathcal L}_{abcd} = \frac{\mathcal P_{ab}}{{\sqrt{M_aM_b}}}\sum_\mu \left[e_\mu^a e_\mu^b\frac{\partial\ln a_\mu}{\partial\Phi_{cd}} + \frac 1 2\frac{\partial (e_\mu^a e_\mu^b)}{\partial\Phi_{cd}}\right]a_\mu^2.
\eeq
In the minimum the only non zero term of the hessian matrix is given by:
\beq
\left.\frac{\partial^2 \fscha}{\partial\Phi_{ab}\partial\Phi_{cd}}\right|_{ \mathbf \Phi = \mathbf K} = -  {\mathcal L}_{abcd},
\label{eq:h1}
\eeq
\beq
\frac{\partial^2\fscha}{\partial\Phi_{ab}\partial\Phi_{cd}} = - \frac{\mathcal P_{ab}}{\sqrt{M_aM_b}}\sum_\mu\left[a_\mu e_\mu^a e_\mu^b\frac{\partial a_\mu}{\partial \Phi_{cd}} + \frac 1 2a_\mu^2 \frac{\partial( e_\mu^a e_\mu^b)}{\partial\Phi_{cd}}\right].
\eeq
Let us start with the term inside the square brackets.
The derivative of $a_\mu$ can be obtained with the chain rule:
\beq
\frac{\partial a_\mu}{\partial \Phi_{cd}} = \frac{\partial a_\mu}{\partial\omega_\mu} \frac{\partial\omega_\mu}{\partial \Phi_{cd}} =  
\frac{\mathcal P_{cd}}{2\omega_\mu}  \frac{e_\mu^c e_\mu^d}{\sqrt{M_cM_d}}
\frac{\partial a_\mu}{\partial\omega_\mu}.
\label{a:phi:diff}
\eeq
The  derivative of the polarization versors can be computed with first order perturbation theory:
\beq
\frac{\partial(e_\mu^a e_\mu^b)}{\partial\Phi_{cd}} =
e_\mu^a \frac{\partial e_\mu^b}{\partial\Phi_{cd}} + e_\mu^b\frac{\partial e_\mu^a}{\partial\Phi_{cd}} 
  = \frac{\mathcal P_{cd}}{\sqrt{M_cM_d}}\sum_\nu^{\nu\neq\mu}
\frac{\left(e_\mu^a e_\nu^b + e_\mu^b e_\nu^a\right)\left(e_\nu^c e_\mu^d + e_\mu^c e_\nu^d\right)}{2(\omega_\mu^2 - \omega_\nu^2)}.
\eeq
We have a complete expression for the hessian matrix:
\beq
\frac{\partial^2 \fscha}{\partial\Phi_{ab}\partial\Phi_{cd}} = -\frac{\mathcal P_{ab}\mathcal P_{cd}}{\sqrt{M_aM_bM_cM_d}} 
\Bigg[\sum_\mu \frac{e_\mu^a e_\mu^b e_\mu^c e_\mu^d}{4\omega_\mu}\frac{\partial a_\mu^2}{\partial\omega_\mu}  +
 \sum_{\mu\nu}^{\mu\neq\nu}\frac{e_\mu^a e_\nu^b( e_\mu^c\varepsilon_\nu^d +  e_\nu^c e_\mu^d)}{4} \left(\frac{a_\mu^2}{\omega_\mu^2 - \omega_\nu^2} + \frac{a_\nu^2}{\omega_\nu^2 - \omega_\mu^2}\right)\Bigg].
\eeq
We can use the bosonic occupation number and write $a_\mu$ as a function of $n_\mu$:
\begin{subequations}
\beq
\label{a:boson}
a_\mu = \sqrt{\frac{\hbar}{\omega_\mu}\left[n_\mu(\beta) + \frac 1 2\right]},
\eeq
\beq
\label{a:deriv:boson}
\frac{a_\mu}{2\omega_\mu}\frac{\partial a_\mu}{\partial \omega_\mu}= -\frac{\hbar}{8\omega_\mu^3}\left(2n_\mu + 1 + 2\beta\hbar\omega_\mu n_\mu^2 +  2\beta\hbar\omega n_\mu\right).
\eeq
\end{subequations}
% \beq
% \frac{1}{4\omega_\mu}\frac{\partial a_\mu^2}{\partial\omega_\mu} = - \frac{\hbar}{8\omega_\mu^3}\left(2n_\mu + 1 + 2 \beta\hbar\omega_\mu n_\mu^2 + 2\beta\hbar\omega_\mu n_\mu\right)
% \eeq
% \beq
% a_\mu^2 = \frac{\hbar}{2\omega_\mu} (2 n_\mu + 1)
% \eeq
Therefore we have:
\begin{align}
\frac{\partial^2 \fscha}{\partial\Phi_{ab}\partial\Phi_{cd}} = \frac{ \hbar\,\mathcal P_{ab}\mathcal P_{cd}}{\sqrt{M_aM_bM_cM_d}} &
\Bigg[\sum_\mu e_\mu^a e_\mu^b e_\mu^c e_\mu^d\frac{2n_\mu + 1 + 2\beta\hbar\omega_\mu n_\mu^2 + 2 \beta\hbar\omega_\mu n_\mu}{8\omega_\mu^3} +\nonumber\\
& - \sum_{\mu\nu}^{\mu\neq\nu}\frac{e_\mu^a e_\nu^b(e_\mu^c e_\nu^d + e_\nu^c e_\mu^d)}{8(\omega_\mu^2 - \omega_\nu^2)}\left(\frac{2n_\mu + 1}{\omega_\mu} - \frac{2\omega_\nu + 1}{\omega_\nu}\right)\Bigg]\label{eq:a:he:1}
\end{align}
It is clear from \eqname~\eqref{eq:a:he:1} that a $\bm\Lambda$ matrix can be introduced so that:
\beq
\frac{\partial^2 \fscha}{\partial\Phi_{ab}\partial \Phi_{cd}} = \frac 1 2\mathcal P_{ab}\mathcal P_{cd} \sum_{\mu\nu} \left(\Lambda_{\mu\nu}^{abcd} + \Lambda_{\mu\nu}^{abdc}\right),
\eeq
where
\begin{subequations}
\beq
\Lambda_{\mu\mu}^{abcd} = \frac{\hbar e_\mu^a e_\mu^b e_\mu^c e_\mu^d}{\sqrt{M_aM_bM_cM_d}} \cdot \frac{2n_\mu + 1 + 2 \beta\hbar\omega_\mu n_\mu^2 + 2 \beta\hbar\omega_\mu n_\mu}{8\omega_\mu^3},
\eeq
\beq
\Lambda_{\mu\nu}^{abcd} = - \frac{\hbar }{\sqrt{M_aM_bM_cM_d}} \frac{e_\mu^a e_\nu^b e_\mu^c e_\nu^d}{(\omega_\mu - \omega_\nu)(\omega_\mu + \omega_\nu)} \frac{2n_\mu\omega_\nu - 2\omega_\mu n_\nu + \omega_\nu - \omega_\mu}{4\omega_\mu\omega_\nu}.
\eeq
\end{subequations}
To conclude the proof it is sufficient to show that the $\bm\Lambda$ matrix of \eqname~\eqref{eq:hessian} is equal to:
\beq
\Lambda^{abcd} = \sum_{\mu\nu} \Lambda_{\mu\nu}^{abcd}.
\eeq
First, we introduce an auxiliary function $f(\omega_\mu, \omega_\nu)$ as
\beq
\label{eq:f:diff}
f(\omega_\mu, \omega_\nu) = \frac{2\omega_\nu n_\mu - 2 \omega_\mu n_\nu + \omega_\nu - \omega_\mu}{4\omega_\mu\omega_\nu(\omega_\mu + \omega_\nu)(\omega_\mu - \omega_\nu)}
= -\frac{1}{4\omega_\mu\omega_\nu}\left[\frac{n_\mu + n_\nu + 1}{\omega_\mu + \omega_\nu} - \frac{n_\mu - n_\nu}{\omega_\mu - \omega_\nu}\right].
\eeq
In the limit $\omega_\nu \rightarrow\omega_\mu$ we get:
\beq
f(\omega_\mu) = \lim_{\omega_\nu \rightarrow\omega_\mu} f(\omega_\mu, \omega_\nu) = - \frac{2n_\mu + 1 + 2\hbar\beta\omega_\mu n_\mu^2 + 2\hbar\beta n_\mu\omega_\mu}{8\omega_\mu^3}
\eeq
\beq
f(\omega_\mu) =-\frac{1}{4\omega_\mu^2}\left[\frac{2n_\mu + 1}{2\omega_\mu} - \frac{\partial n}{\partial\omega} \right] .
\label{eq:f:eq}
\eeq
So $\Lambda_{\mu\mu}^{abcd}$ is obtained as the continue limit of $\Lambda_{\mu\nu}^{abcd}$ when $\mu\rightarrow\nu$:
\beq
\Lambda_{\mu\nu}^{abcd} = -\frac{\hbar e_\mu^a e_\mu^b e_\mu^c e_\mu^d}{\sqrt{M_aM_bM_cM_d}} f(\omega_\mu, \omega_\nu),\qquad
\Lambda_{\mu\mu}^{abcd} = -\frac{\hbar e_\mu^a e_\mu^b e_\mu^c e_\mu^d}{\sqrt{M_aM_bM_cM_d}} f(\omega_\mu).
\eeq
Substituting \eqname~\eqref{eq:f:diff} and~\eqref{eq:f:eq} we finally get:
\beq
\Lambda^{abcd}_{\mu\nu} = \frac{\hbar}{4\omega_\mu\omega_\nu}\frac{e_\mu^{a} e_\nu^{b} e_\mu^c e_\nu^d}{\sqrt{M_aM_bM_cM_d}} \times
\left\{\begin{array}{lr}
\displaystyle
\frac{n_\mu + n_\nu + 1}{\omega_\mu + \omega_\nu} - \frac{n_\mu - n_\nu}{\omega_\mu - \omega_\nu} & \omega_\mu\neq\omega_\nu \\
\\\displaystyle
\frac{2n_\mu + 1}{2\omega_\mu} - \frac{\partial n_{\mu}}{\partial\omega_\mu} & \omega_\mu = \omega_\nu
\end{array}\right..
\eeq

\section{QHA in the SCHA framework}
\label{app:QHA:vs:SCHA}
The quasi-harmonic approximation (QHA) can be reformulated in the SCHA framework in order to understand differences between the two approaches.
The SCHA free energy is:
\beq
\fscha = F_{\mathbf\Phi} + \braket{V - \mathcal V_{\vec{\mathcal R}, \mathbf\Phi}}_\rscha .
\label{eq:f:scha}
\eeq
If the system is perfectly harmonic, then the minimum of the free energy is found when $\mathcal V_{\vec R_0, \mathbf\Phi_0} = V$, and we get the QHA free energy:
\beq
F_{QHA} = \mathcal F = F_{\mathbf\Phi_0} + V(\vec R_0),\qquad {\Phi_0}_{\alpha\beta} = \left.\frac{\partial^2V}{\partial R_\alpha \partial R_\beta}\right|_{\vec R = \vec {R_0}}.
\eeq
where $\vec{R_0}$ is the minimum of the BO energy surface.
So QHA is equivalent to SCHA for any harmonic potential. If the system is anharmonic, QHA approximates the potential as the second order Taylor expansion around
the equilibrium position. This makes the QHA theory not a self-consistent approach, but a series expansion of the real potential.

If the atomic position coordinates relaxation is allowed, as introduced by Lazzeri and de~Gironcoli\cite{Lazzeri_1998,Lazzeri_2002}, then the QHA free energy becomes:
\beq
\mathcal F_{QHA}(\vec R_c) = F_{\tilde{\mathbf\Phi}(\vec R_c)} + V(\vec R_c),\qquad 
\tilde\Phi_{\alpha\beta}(\vec R_c) = \left.\frac{\partial^2V}{\partial R_\alpha \partial R_\beta}\right|_{\vec R = \vec {R_c}}.
\eeq
This is equivalent to SCHA (\eqname~\ref{eq:f:scha}) keeping $\Phi$ fixed to the harmonic dynamical matrix and neglecting the contribution arising from
$\braket{V - \mathcal V}_{\rho_{\vec R_c, \tilde {\mathbf\Phi}(\vec R_c)}}$. The anharmonicity is taken into account by the fact that the harmonic dynamical matrix is a function of the atomic positions.
This approximation is equivalent to neglecting all the even (from the fourth order) contribution in the BO surface Taylor expansion around the $\vec R_c$ that minimizes $\mathcal F_{QHA}$. 
In this case the average $\braket{V - \mathcal V}_{\rho_{\tilde{\mathbf\Phi}(\vec R_c)}}$ is equal to zero, and
the harmonic dynamical matrix is the one that minimizes the SCHA free energy (\eqname~\ref{eq:gradient} is exactly zero). 
If only odd anharmonicities are present in the system (i.e. they dominate in the region of the quantum and thermal fluctuations), the QHA relaxed free energy coincides with the SCHA. The SCHA, therefore, is a natural extension to the relaxed QHA
that assures the self-consistency of the theory for any kind of anharmonicity by explicitly including the average $\braket{V - \mathcal V}_\rscha$ in the free energy.
Indeed, the SSCHA algorithm is much more efficient than the  QHA relaxation, since it requires only to compute energies and forces, while the QHA relaxation requires third order
derivatives of the energy, and the application of the $2n + 1$ theorem for each minimization step\cite{Lazzeri_2002}.

\end{widetext}
\bibliography{biblio}
\bibliographystyle{ieeetr}

\end{document}